\documentclass[twocolumn,showpacs,aps,prb]{revtex4-1}
\usepackage{bm}
\usepackage{amssymb,amsmath}

\usepackage[dvips]{graphicx}

\begin{document}

\title{Monte Carlo simulations of the disordered three-color quantum Ashkin-Teller chain}

\author{Ahmed K. Ibrahim}

\author{Thomas Vojta}
\affiliation{Department of Physics, Missouri University of Science and Technology, Rolla, Missouri 65409, USA}

\begin{abstract}
We investigate the zero-temperature quantum phase transitions of the disordered
three-color quantum Ashkin-Teller spin chain
by means of large-scale Monte Carlo simulations. We find that the first-order phase transitions of the clean
system are rounded by the quenched disorder. For weak inter-color coupling, the resulting emergent quantum critical
point between the paramagnetic phase and the magnetically ordered Baxter phase is of infinite-randomness type and
belongs to the universality class of the random transverse-field
Ising model, as predicted by recent strong-disorder renormalization group calculations.
We also find evidence for unconventional critical behavior in the case of strong inter-color coupling,
even though an unequivocal determination of the universality class is beyond our numerical capabilities.
We compare our results to earlier simulations, and we discuss implications for the classification of
phase transitions in the presence of disorder.
\end{abstract}

\date{\today}

\maketitle
\section{Introduction}
\label{Intro}

Zero-temperature quantum phase transitions can be classified into continuous or first-order
just as classical thermal phase transitions. First-order quantum phase transitions have gained
considerable attention recently, not only because of their fundamental interest but also
because experimentally important transitions turn from being continuous at higher
temperatures to first-order at lower temperatures. A prominent example of this behavior
is the itinerant ferromagnetic transition.\cite{BelitzKirkpatrickVojta97,*BelitzKirkpatrickVojta99,BelitzKirkpatrickVojta05}
(For a recent review of metallic quantum ferromagnets see Ref.\
\onlinecite{BrandoBelitzGroscheKirkpatrick16}.)

As real materials always contain a certain amount vacancies, impurities, or other
defects, understanding the influence of such quenched disorder is of both conceptual
and practical importance. Theoretical research on continuous quantum phase transitions in
the presence of disorder has predicted a number of exotic phenomena such as
infinite-randomness critical points,
\cite{Fisher92,*Fisher95,MMHF00,HoyosKotabageVojta07,*VojtaKotabageHoyos09}
quantum Griffiths phases, \cite{ThillHuse95,RiegerYoung96,*YoungRieger96} and
smeared phase transitions. \cite{Vojta03a,*HoyosVojta08}
More recently, several of these phenomena have been observed in experiments.
\cite{GYMBHCHD07,*GYMBHCHD10a,Westerkampetal09,UbaidKassisVojtaSchroeder10,Demkoetal12}
A classification of strong-disorder effects was developed in Ref.\ \onlinecite{VojtaSchmalian05} and refined
in Ref.\ \onlinecite{VojtaHoyos14}, see also
Refs.\ \onlinecite{Vojta06,*Vojta10,*Vojta14} for reviews.

In contrast, less is known about first-order quantum phase transitions in the
presence of disorder. Greenblatt et al. \cite{GreenblattAizenmanLebowitz09,*AizenmanGreenblattLebowitz12}
proved a quantum version of the classical Aizenman-Wehr theorem \cite{ImryWortis79,HuiBerker89,AizenmanWehr89}
that states that first-order phase transitions cannot exist in disordered systems
in $d\le 2$ space dimensions. (If the disorder breaks a continuous symmetry, the marginal dimension is
$d=4$.) This agrees with a few available explicit results:
Senthil and Majumdar \cite{SenthilMajumdar96} predicted that quenched randomness turns the
first-order quantum phase transitions of the quantum Potts and
clock chains into infinite-randomness critical points in the random transverse-field Ising universality
class. The same was found by Goswami et al.
\cite{GoswamiSchwabChakravarty08} for the disordered $N$-color one-dimensional quantum Ashkin-Teller model
\cite{Fradkin84,*Shankar85} in the weak-coupling regime (weak interactions between the colors).
In the strong-coupling regime, the critical point between the para\-magnetic and Baxter phases is still of infinite-randomness type,
but it is predicted to be in a different universality class. \cite{HrahshehHoyosVojta12,Barghathietal14}

All these results were obtained using versions of the strong-disorder renormalization group
\cite{IgloiMonthus05} which becomes controlled in the limit of infinitely strong disorder. It is
therefore highly desirable to verify that the predictions also hold for realistic, weakly or moderately
disordered systems. A recent Monte Carlo study of the quantum Ashkin-Teller model
\cite{BellafardChakravarty16} provided evidence for the activated scaling expected at an
infinite-randomness critical point. However, the authors could not verify the predicted
random transverse-field Ising universality class and suggested that the discrepancy stems,
perhaps, from the first-order origin of this transition.

To shed some light onto this question, we map the disordered three-color quantum Ashkin-Teller
chain onto a $(1+1)$ dimensional classical Hamiltonian with columnar disorder. We investigate this classical
model by means of large-scale Monte Carlo simulations for systems with up to 3.6 million lattice sites
(10.8 million spins). In the weak-coupling regime, we find universal critical behavior in the
random transverse-field Ising universality class, as predicted by the strong-disorder renormalization
group. We also perform exploratory simulations in the strong coupling regime that establish the
phase diagram and confirm unconventional activated dynamical scaling.
However, because the efficient cluster Monte Carlo algorithms we use in the weak-coupling
regime are not valid for strong coupling,
we can not quantitatively verify the distinct critical behavior predicted in Refs.\
\onlinecite{HrahshehHoyosVojta12,Barghathietal14}.

The rest of the paper is organized as follows. In Sec.\ \ref{sec:Model}, we introduce the quantum
Ashkin-Teller chain and the mapping onto a classical Hamiltonian. We also summarize the predictions of
the strong-disorder renormalization group calculations. Section \ref{sec:MC} is devoted
to the Monte Carlo simulations and their results. We conclude in Sec.\ \ref{sec:Conclusions}.


\section{Model and theory}
\label{sec:Model}
\subsection{Quantum Ashkin-Teller chain}
\label{subsec:QAT_chain}
The $N$-color quantum Ashkin-Teller chain \cite{GrestWidom81,Fradkin84,*Shankar85}
is a generalization of the original model suggested by Ashkin and Teller many
decades ago. \cite{AshkinTeller43} It is made up of $N$ coupled identical
transverse-field Ising chains each containing $L$ spins. The quantum Hamiltonian can be
expressed as
\begin{eqnarray}
\label{eq:HAT}
 H&=&-\sum_{\alpha=1}^N\sum_{i=1}^L{\left ( J_i \sigma_{\alpha,i}^z \sigma_{\alpha,i+1}^z
+ h_i \sigma_{\alpha,i}^x \right )}\\
&&-\sum_{\alpha<\beta}^N\sum_{i=1}^L{\left (K_i \sigma_{\alpha,i}^z
\sigma_{\alpha,i+1}^z \sigma_{\beta,i}^z \sigma_{\beta,i+1}^z
+ g_i \sigma_{\alpha,i}^x \sigma_{\beta,i}^x\right )}~~.
\end{eqnarray}
Here,  $\sigma^x$ and $\sigma^z$ are Pauli matrices describing the spin degrees of freedom. $i$ denotes the lattice sites while $\alpha$ and $\beta$ are color indices.
The ratios $\epsilon_{h,i}=g_i/h_i$ and $\epsilon_{J,i}=K_i/J_i$ characterize the strengths of the inter-color couplings.
In the following, we are interested in the case of positive interactions $J_i$, $K_i$ and fields $h_i$, $g_i$.
Besides its fundamental interest, different versions of the Ashkin-Teller model have been used to
describe absorbed atoms on surfaces \cite{Baketal85}, organic magnets, current loops in
high-temperature superconductors \cite{AjiVarma07,AjiVarma09}, as well as the elastic response of DNA
molecules.\cite{ChangWangZheng08}

In the clean quantum Ashkin-Teller chain, the interactions and coupling strengths are uniform in space,
$J_i\equiv J$, $K_i\equiv K$, $h_i\equiv h$, and $g_i\equiv g$. The bulk phases of this model are easily understood qualitatively.
In the weak-coupling regime $\epsilon_{J}, \epsilon_{h} \ll 1$, the system is in the paramagnetic phase
if the transverse fields are larger than the interactions, $h \gg J$.
For $h \ll J$, the system is in the ordered (Baxter) phase in which each color orders ferromagnetically but
the relative orientation of different colors is arbitrary.  Another phase, the so-called product phase, can appear
in the strong coupling regime $\epsilon_{J}, \epsilon_{h} \gg 1$.  In this phase, products
$\sigma_{\alpha,i}^z \sigma_{\beta,i}^z$ of two spins of different colors develop long-range order
while the spins $\sigma_{\alpha,i}^z$ themselves remain disordered. For at least three colors, the direct quantum phase transition
between the paramagnetic and Baxter phases is known to be of first-order. \cite{GrestWidom81,Fradkin84,*Shankar85,Ceccatto91}
The quantum Ashkin-Teller chain is therefore a paradigmatic model for studying the effects of disorder on a
first-order quantum phase transition.

Note that the form of the Hamiltonian (\ref{eq:HAT}) is invariant under the duality transformation
$\sigma_{\alpha,i}^z \sigma_{\alpha,i+1}^z \to \tilde \sigma_{\alpha,i}^x$,
$\sigma_{\alpha,i}^x \to \tilde \sigma_{\alpha,i}^z \tilde \sigma_{\alpha,i+1}^z$,
 $J_i\rightleftarrows h_i$, and $\epsilon_{J,i}\rightleftarrows\epsilon_{h,i}$,
where $\tilde \sigma_{\alpha,i}^x$ and $\tilde \sigma_{\alpha,i}^z$ are the dual Pauli matrices
\cite{Baxter_book82}. Self-duality therefore requires that a direct transition between
the paramagnetic and Baxter phases (for $\epsilon_h=\epsilon_J$) must occur
exactly at $h=J$.

\subsection{Renormalization group predictions}
\label{subsec:RG}

We now briefly summarize the results of several strong-disorder
renormalization group calculations for the $N$-color random quantum Ashkin-Teller chain.
 Goswami et al.\ \cite{GoswamiSchwabChakravarty08} analyzed the weak-coupling regime
and found that the inter-color coupling strengths $\epsilon_{J,i}, \epsilon_{h,i}$ renormalize to zero,
and the renormalization group flow  becomes asymptotically identical to that of the one-dimensional random transverse-field Ising
model.\cite{Fisher92,*Fisher95}
More specifically, this happens if all initial (bare) $\epsilon_{J,i}$ and $\epsilon_{h,i}$ are smaller than
a critical value
\begin{equation}
\epsilon_c(N) = \frac {2N-5}{2N-2} + \sqrt{\left(\frac {2N-5}{2N-2}\right)^2 + \frac{2}{N-1}}~.
\label{eq:epsilon_c}
\end{equation}
(For three colors, $\epsilon_c\approx 1.281$.)
In the weak-coupling regime, the strong disorder renormalization group thus predicts that the first-order quantum phase transition
of the clean chain is rounded to a continuous one, with infinite-randomness critical behavior in
the random transverse-field Ising universality class.\cite{Fisher92,*Fisher95}

The strong-coupling regime of the random quantum Ashkin-Teller chain was studied in Refs.\
\onlinecite{HrahshehHoyosVojta12,HHNV14,Barghathietal14}. Using a different implementation
of the strong-disorder renormalization group, these papers demonstrated 
that the inter-color coupling strengths $\epsilon_{J,i}$ and $\epsilon_{h,i}$ renormalize to infinity
if their initial (bare) values are larger than $\epsilon_c$. This implies that the
four-spin interactions and the two-spin field terms in the Hamiltonian dominate the behavior of the
system.

If $\epsilon_{J,i} = \epsilon_{h,i}$, the model is self-dual at the critical point.
In this case and for at least three colors, there is still a direct transition between the paramagnetic and Baxter
phases, i.e., spins and products order at the same point. This transition occurs at $J_{\rm typ}=h_{\rm typ}$
where $J_{\rm typ}$ and $h_{\rm typ}$ refer to the typical values (geometric means) of the
random interactions and fields. The critical behavior of this transition is of infinite randomness type
but it is not in the random transverse Ising universality class because products and spins
both contribute to observables. \cite{HrahshehHoyosVojta12,Barghathietal14}
In the general case, $\epsilon_{J,i} \ne \epsilon_{h,i}$, a product phase can appear
between the paramagnetic and Baxter phases (this also happens for two colors, even in the self-dual case).
\cite{HHNV14} The phase transition between the paramagnetic and product phases
as well as the transition between the product and Baxter phases are both expected to belong
to the random transverse-field Ising universality class.

\subsection{Quantum-to-classical mapping}
\label{subsec:Mapping}

To test the renormalization group predictions by Monte Carlo simulations, we now map the random
quantum Ashkin-Teller chain onto a (1+1)-dimensional classical Ashkin-Teller model.
This can be done using standard methods, e.g., by writing the partition
function as a Feynman path integral in imaginary time (see also Ref.\ \onlinecite{Sachdev_book99}).
The resulting classical Hamiltonian reads:
\begin{eqnarray}
H_{cl}&=& -\sum_{\alpha, i,t} \left ( J_i^{(s)} S^\alpha_{i,t} S^\alpha_{i+1,t} + J_i^{(t)} S^\alpha_{i,t} S^\alpha_{i,t+1} \right)\nonumber \\
      & & -\sum_{\alpha<\beta, i,t} \left ( \epsilon_i^{(s)} J_i^{(s)} S^\alpha_{i,t} S^\alpha_{i+1,t}S^\beta_{i,t} S^\beta_{i+1,t} \right) \nonumber \\
      & & -\sum_{\alpha<\beta, i,t} \left ( \epsilon_i^{(t)} J_i^{(t)} S^\alpha_{i,t} S^\alpha_{i,t+1}S^\beta_{i,t} S^\beta_{i,t+1} \right )~.
\label{eq:Hcl}
\end{eqnarray}
Here, $S^\alpha_{i,t}=\pm 1$ is a classical Ising spin of color $\alpha$ at position $i$ in
space and $t$ in (imaginary) time. The classical interactions
$J_i^{(s)}$, $J_i^{(t)}$ and coupling constants $\epsilon_i^{(s)}$, $\epsilon_i^{(t)}$ as well as
the classical temperature $T$
are determined by the parameters of
the original quantum Hamiltonian (\ref{eq:HAT}). (The classical temperature $T$ does not equal the
physical temperature of the quantum system (\ref{eq:HAT}) which is encoded in the system size
$L_t$ in time direction.)
As we are interested in the critical behavior
which is expected to be universal, the precise values of
$J_i^{(s)}$, $J_i^{(t)}$, $\epsilon_i^{(s)}$, and $\epsilon_i^{(t)}$ are not important and can be chosen
for computational convenience (see Sec.\ \ref{sec:MC}).

We note that the quantum-to-classical mapping generates further terms in the classical
Hamiltonian in addition to those shown in (\ref{eq:Hcl}). These extra terms contain
higher products of up to $N$ colors. Neglecting them does not change the critical behavior,
but it destroys the self-duality of the Hamiltonian.

\section{Monte Carlo simulations}
\label{sec:MC}
\subsection{Overview}
\label{subsec:Overview}

We perform large-scale Monte Carlo simulations of the classical Hamiltonian (\ref{eq:Hcl})
for the case of $N=3$ colors by employing an Ising embedding method similar that used
in Ref.\ \onlinecite{WisemanDomany95}.
It can be understood as follows. If we fix the values of all spins with color $\alpha\ne1$,
the Hamiltonian (\ref{eq:Hcl}) acts as an (1+1)-dimensional Ising model for the spins
$S^{(1)}_{i,t}$ with effective interaction
$J_{ij}^\textrm{eff} = J + \epsilon J ( S_i^{(2)}S_j^{(2)} + S_i^{(3)}S_j^{(3)})$.
This embedded Ising model can be simulated by means of any Ising Monte Carlo algorithm.
We use a combination of the efficient Swendsen-Wang multicluster algorithm \cite{SwendsenWang87} and
the Wolff single cluster alorithm.\cite{Wolff89} Analogous embedded Ising models can be
constructed for the spins $S^{(2)}_{i,t}$ and $S^{(3)}_{i,t}$, and by performing cluster
updates for all three embedded Ising models we arrive at a valid and efficient
algorithm for the Ashkin-Teller model.

The Swendsen-Wang and Wolff cluster algorithms require all interactions to be
nonnegative, $J^\textrm{eff} \ge 0$.
\footnote{Generalizations of the Swendsen-Wang and Wolff algorithms exist
for systems with competing interaction, but they turn out be much less efficient
\cite{KesslerBretz90,Liang92}}
This is only guaranteed if the coupling constant
$\epsilon$ does not exceed $1/(N-1)=1/2$. For larger $\epsilon$, we perform
exploratory simulations using the less efficient Metropolis algorithm\cite{MRRT53}
as well as the Wang-Landau method. \cite{WangLandau01}

By means of these algorithms, we simulate systems with linear sizes
$L=10$ to 60 in space direction and $L_t=2$ to 60,000 in (imaginary) time direction,
using periodic boundary conditions.
The largest system had 3.6 million lattice sites, i.e., 10.8 million spins.
To implement the quenched disorder, we consider $J_i^{(s)}$ and $J_i^{(t)}$ to be independent
random variables drawn from a binary probability distribution
\begin{equation}
W(J) = c\delta(J-J_h) +(1-c)\delta(J-J_l)
\label{eq:P(J)}
\end{equation}
where $c$ is the concentration of the higher value $J_h$ of the interaction while $1-c$ is the concentration of
the lower value $J_l$. The couplings are uniform, $\epsilon_{i}^{(s)}=\epsilon_{i}^{(t)}=\epsilon$.
As $J_i^{(s)}$ and $J_i^{(t)}$ only depend on the space coordinate $i$ but not on the time coordinate $t$, the resulting
disorder is columnar, i.e., perfectly correlated in the time direction.
In the simulations, we use $J_h=1$, $J_l=0.25$, and $c=0.5$ while $\epsilon$ takes
values between 0 and 5. All observables are averaged over 10,000 to 40,000 disorder configurations,
unless otherwise noted.

When using cluster algorithms ($\epsilon \le 0.5$),
we equilibrate each sample using 100 full Monte Carlo sweeps. Each full sweep
is made up of a Wolff sweep for each color (consisting of a number of
single-cluster flips such that the total number of flipped spins equals
the number of lattice sites) and a Swendsen-Wang sweep for each color. The
Swendsen-Wang sweep aims at equilibrating small clusters of weakly coupled sites
that may be missed by the Wolff algorithm.
The actual equilibration is significantly faster than 100 sweeps. \cite{ZWNHV15}
The measurement period
consists of another 100 full Monte Carlo sweeps with a measurement taken after each sweep.
To deal with biases introduced by using such short measurement periods, we
employ improved estimators. \cite{ZWNHV15} Simulations for $\epsilon>0.5$
that use the Metropolis and Wang-Landau methods require much longer runs, details
will be discussed below.

During the simulation runs, we measure the following observables: energy,
specific heat, total magnetization
\begin{equation}
m=\frac 1 {3 L L_t} \sum_{\alpha} \left| \sum_{i,t} S_{i,t}^\alpha \right|
\label{eq:mag}
\end{equation}
and its susceptibility $\chi_m$. A particularly useful quantity for
the finite-size scaling analysis is the Binder cumulant
\begin{equation}
g_\textrm{av}= \left[ 1-\frac {\langle m^4 \rangle}{3\langle m^2\rangle^2 }  \right]_\textrm{dis}
\label{eq:Binder}
\end{equation}
where $\langle \ldots \rangle$ denotes the thermodynamic (Monte Carlo) average
and $[\ldots]_\textrm{dis}$ is the disorder average. In addition, we also
measure the product order parameter
\begin{equation}
p=\frac 1 {3 L L_t} \sum_{\alpha<\beta} \left| \sum_{i,t} S_{i,t}^\alpha S_{i,t}^\beta \right|~,
\label{eq:prod}
\end{equation}
the corresponding product susceptibility $\chi_p$, and the product Binder cumulant $g_p$.

\begin{figure}
\includegraphics[width=8cm]{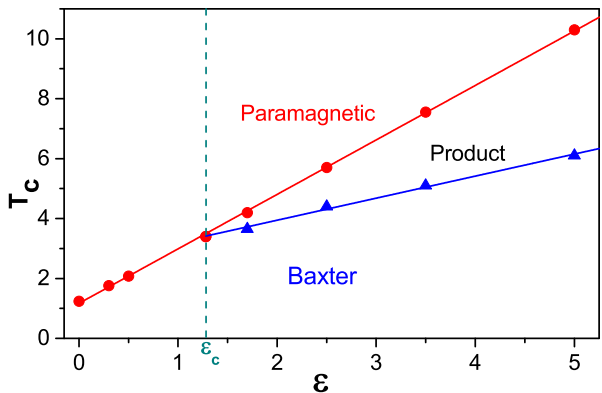}
\caption{(Color online) Phase diagram of the classical Hamiltonian (\ref{eq:Hcl})
        for $N=3$ colors and disorder distribution
        (\ref{eq:P(J)}) with $J_h=1$, $J_l=0.25$, and $c=0.5$. The dots
        and triangles mark the numerically determined transitions between the
        Baxter, product, and paramagnetic phases. The solid lines are guides to the eye
        only. The dashed line marks $\epsilon_c=1.281$ [see Eq.\  (\ref{eq:epsilon_c})]
        which separates the weak and strong coupling regimes in the strong-disorder
        renormalization group calculations. }
\label{fig:pd}
\end{figure}

The phase diagram of the classical Hamiltonian (\ref{eq:Hcl}) resulting
from these simulations is shown in Fig.\ \ref{fig:pd}.
In the weak-coupling regime, $\epsilon<\epsilon_c$, we find a direct transition between
the magnetically ordered Baxter phase at low temperatures and the paramagnetic
high-temperature phase. For strong coupling, $\epsilon>\epsilon_c$, these two phases
are separated by a product phase.
In the following, we study the critical behaviors of the transitions
separating these phases in detail,  and we compare them to the renormalization group
predictions.

\subsection{Weak coupling regime}
\label{subsec:MC_weak}

In the weak-coupling regime, $\epsilon<\epsilon_c$, we perform simulations for
coupling strengths $\epsilon=0$, 0.3 and 0.5 employing the Wolff and Swendsen-Wang
cluster algorithms as discussed above.
Because the disorder breaks the symmetry between the space and (imaginary) time directions
in the Hamiltonian (\ref{eq:Hcl}), the finite-size scaling analysis of the data to
find the critical exponents becomes more complicated. This is caused by the fact that
the system sizes $L$ and $L_t$ in the space and time directions are expected to have
different scaling behavior. Thus, the correct aspect ratios $L_t/L$ of the samples
to be used in the simulations are not known a priori.

To overcome this problem we follow the iterative method employed
in Refs.\ \onlinecite{GuoBhattHuse94,RiegerYoung94,SknepnekVojtaVojta04,*VojtaSknepnek06,Vojtaetal16}
which is based on the Binder cumulant. As the renormalization group calculations
predict infinite-randomness criticality with activated dynamical scaling, the scaling
form of the Binder cumulant (which has scale dimension 0) reads
\begin{equation}
g_\textrm{av}(r,L,L_t) = X_g(rL^{1/\nu},\ln(L_t/L_{t}^0)/L^\psi)~.
\label{eq:Binderscaling}
\end{equation}
Here $r=(T-Tc)/T_c$ denotes the distance from criticality, $X_g$ is a scaling function, and
$\psi$ and $\nu$ refer to the tunneling and correlation length critical exponents. $L_{t}^0$
is a microscopic reference scale. (For conventional power-law scaling, the second argument of the scaling function would read
$L_t/L^z$ with $z$ being the dynamical exponent.)
For fixed $L$, $g_{\rm av}$ has a maximum as function of $L_{t}$ at position $L_{t}^{\rm max}$
and value $g_{\rm av}^{\rm max}$. The position of the maximum yields the {\em optimal}
sample shape for which the system sizes $L$ and $L_t$ behave as the correlation lengths $\xi$ and $\xi_t$.
At criticality $L_t$ must thus behave as $\ln(L_{t}^{\rm max}/L_t^0) \sim L^\psi$, fixing the second
argument of the scaling function $X_g$. Consequently, the peak value $g_{\rm av}^{\rm max}$
is independent of $L$ at criticality, and the  $g_{\rm av}$ vs.\ $r$ curves of optimally shaped samples
cross at $T=T_c$. Once the optimal sample shapes are found, finite-size scaling proceeds as usual.
 \cite{Barber_review83,Cardy_book88}

To test our simulation and data analysis technique, we first consider the case $\epsilon=0$
for which the quantum Ashkin-Teller model reduces to three decoupled random transverse-field
Ising chains whose quantum phase transition is well understood.\cite{Fisher92,*Fisher95}
We perform simulations for sizes $L=10$ to 50 and $L_t=2$ to 20000 and find a
critical temperature $T_c\approx 1.24$.
At this temperature, we confirm the activated scaling (\ref{eq:Binderscaling})
of the Binder cumulant with the expected value $\psi=1/2$. We also confirm the scaling
of the magnetization at $T_c$ (for the optimally shaped samples),
$m \sim L^{-\beta/\nu}$ with $\beta=0.382$ and  $\nu=2$.

After this successful test, we now turn to the Ashkin-Teller model proper. We perform two sets of
simulations: (i) $\epsilon=0.5$ using system sizes $L=10$ to 60, $L_t=2$ to 60000
and (ii) $\epsilon=0.3$ with system sizes $L=10$ to 50, $L_t=2$ to 40000.
In each case, we start from a guess for the optimal shapes and find an approximate value
of $T_c$ from the crossing of the  $g_{\rm av}$ vs.\ $T$ curves for different $L$. We then
find the maxima of the $g_{\rm av}$ vs.\ $L_t$ curves at this temperature
which yield improved optimal shapes. After iterating this procedure two or three times,
we obtain $T_c$ and the optimal shapes with reasonable precision.

\begin{figure}
\includegraphics[width=8.2cm]{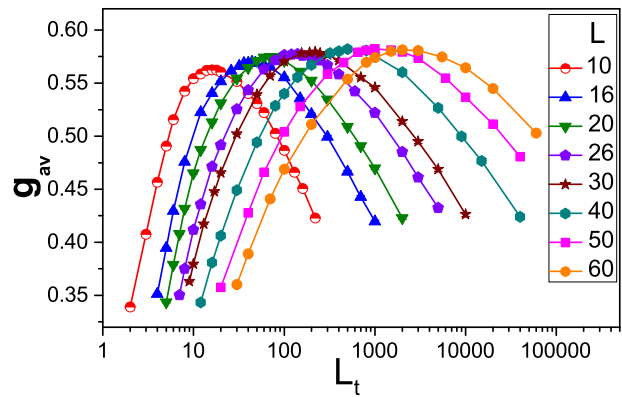}
\caption{(Color online) Binder cumulant $g_{\rm av}$ as a function of $L_t$ for several $L$
        at the critical temperature $T_c=2.08$ for $\epsilon=0.5$. The statistical error of
        $g_{\rm av}$ is smaller than the symbol size.}
\label{fig:gLLteps05}
\end{figure}

\begin{figure}
\includegraphics[width=8.2cm]{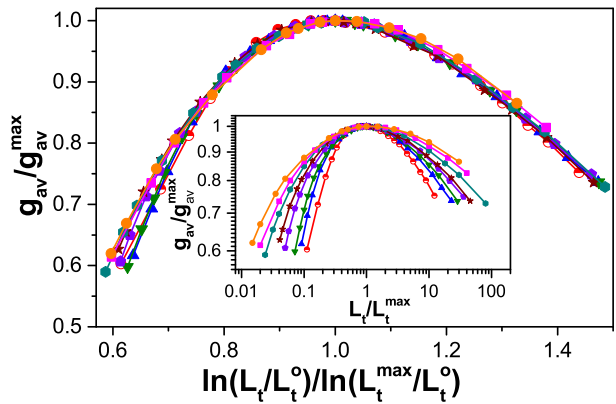}
\caption{(Color online) Scaling plot of the Binder cumulant at $T_c=2.08$ for $\epsilon=0.5$. The symbols are
          the same as in Fig.\ \ref{fig:gLLteps05}
          Main panel: Activated scaling $g_{\rm av}/g_{\rm av}^{\rm max}$ vs.\  $\ln(L_t/L_t^0)/\ln(L_t^{\rm max}/L_t^0)$
          according to Eq.\ (\ref{eq:Binderscaling}). The microscopic scale $L_t^0=0.06$. Inset: Power-law scaling
          $g_{\rm av}/g_{\rm av}^{\rm max}$ vs.\ $L_t/L_t^{\rm max}$.
          }
\label{fig:gLLteps05scaling}
\end{figure}

Figure \ref{fig:gLLteps05}
shows the resulting Binder cumulant $g_\textrm{av}$ for $\epsilon=0.5$
as function of $L_t$ for different $L$ at the approximate critical temperature of $T_c = 2.08(5)$.
As expected at $T_c$, the maxima $g_{\rm av}^{\rm max}$ of these curves are independent
of $L$ (the slightly lower values at the smallest $L$ can be attributed to corrections to scaling).
Moreover, the figure shows that the $g_{\rm av}$ vs.\ $L_t$ domes rapidly become broader
with increasing spatial size $L$, indicating non-power-law scaling. To analyze this quantitatively,
we present a scaling plot of these data in Fig.\ \ref{fig:gLLteps05scaling}.
For conventional power-law dynamical scaling, the curves for different $L$ should
collapse onto each other when plotted as $g_{\rm av}$ vs.\ $L_t/L_t^{\rm max}$.
The inset of Fig.\ \ref{fig:gLLteps05scaling} clearly demonstrates that this is not the case.
In contrast, the Binder cumulant scales well when plotted versus $\ln(L_t/L_t^0)/\ln(L_t^{\rm max}/L_t^0)$
as shown in the main panel of the figure. (Here, we treat the microscopic scale $L_t^0$
as a fit parameter). This behavior is in agreement with the activated
scaling form (\ref{eq:Binderscaling}).

\begin{figure}
\includegraphics[width=8.2cm]{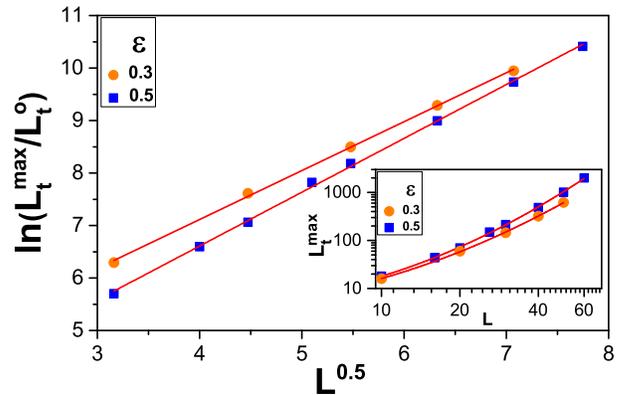}
\caption{(Color online) $\ln(L_t^{\rm max}/L_t^0)$ vs.\ $L^{0.5}$ at criticality for $\epsilon=0.3$ and 0.5.
         The data for $\epsilon=0.3$ are shifted upwards by 0.3 for clarity. The solid lines are linear fits.
         Inset: Double logarithmic plot of $L_t^{\rm max}$ vs.\ $L$.
          }
\label{fig:LLtmax}
\end{figure}

We perform the same analysis for $\epsilon=0.3$ at the approximate
critical temperature of $T_c =1.76(3)$, with analogous results. To verify the value
of the tunneling exponent $\psi$, we now analyze the dependence of
$L_t^{\rm max}$ on $L$. Figure\ \ref{fig:LLtmax} shows that the data for both
$\epsilon=0.3$ and 0.5 can be well fitted with the relation $\ln(L_t^{\rm max}/L_t^0) \sim L^\psi$
with $\psi=1/2$ as predicted by the strong-disorder renormalization group.
The inset of this figure clearly demonstrates that the relation between $L_t^{\rm max}$ and $L$
cannot be described by a power law. We can define, however, an effective (scale-dependent) dynamical exponent
$z_{\rm eff}= d\ln(L_t^{\rm max})/d\ln(L)$. For $\epsilon=0.5$, it increases from about 2
for the smallest system sizes to almost 4 for the largest ones.

\begin{figure}
\includegraphics[width=8.2cm]{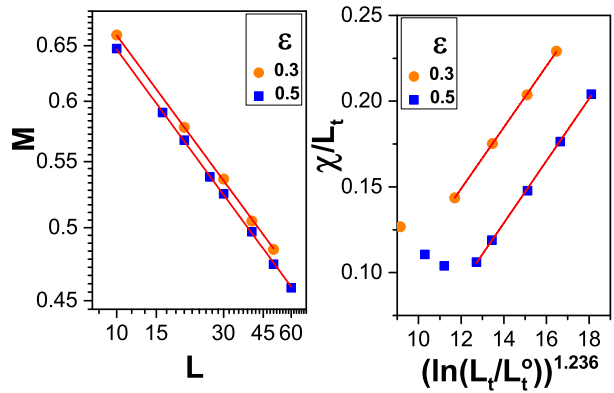}
\caption{(Color online) Left: Double logarithmic plot of $m$ vs.\ $L$ for optimally
         shaped samples at criticality for $\epsilon=0.3$ and 0.5.
         The solid lines are fits to the predicted power-law $m \sim L^{-\beta/\nu}$
         with $\beta/\nu =0.191$. Right: $\chi/L_t$ vs.\ $[\ln(L_t/L_t^0)]^{2\phi-2}$
         for optimally shaped samples at criticality for $\epsilon=0.3$ and 0.5.
         The solid lines are linear fits. The statistical errors of the data in both panels
         are smaller than the symbol size.
          }
\label{fig:magL}
\end{figure}

We now turn to the critical behavior of magnetization and susceptibility.
At the critical temperature, the magnetization of the optimally shaped samples
is predicted to show a power-law dependence on the spatial system size,
$m \sim L^{-\beta/\nu}$ with $\beta=2-\phi\approx 0.382$ and $\nu=2$. Here,
$\phi=(\sqrt{5}+1)/2$ is the golden mean. In the left panel of
Fig.\ \ref{fig:magL}, we therefore
present a double logarithmic plot of $m$ vs.\ $L$ for $\epsilon=0.3$ and
0.5.
The data for both coupling strengths can be fitted well with the predicted power
law. While the magnetization follows a conventional power law dependence on the system
size, the susceptibility is affected by the activated scaling. Its predicted system
size dependence at criticality can be expressed in terms of the temporal size $L_t$
as $\chi \sim L_t [\ln(L_t/L_t^0)]^{2\phi-2}$. We test this prediction
in the right panel of Fig.\ \ref{fig:magL} by plotting $\chi/L_t$ vs.\ $[\ln(L_t/L_t^0)]^{2\phi-2}$
for the optimally shaped samples. As the leading power law is divided out, this plot
provides a sensitive test of the logarithmic corrections. The figure shows that the
susceptibility indeed follows the predicted $L_t$ dependence for system sizes $L>20$.
The deviations for the smaller sizes can likely be attributed to corrections
to scaling stemming from the crossover between the clean first-order phase transition
and the infinite-randomness critical point that governs the asymptotic behavior.
The clean first-order phase transition is stronger for $\epsilon=0.5$ than for 0.3;
accordingly, $\chi$ shows stronger corrections to scaling for $\epsilon=0.5$.

\begin{figure}
\includegraphics[width=8.2cm]{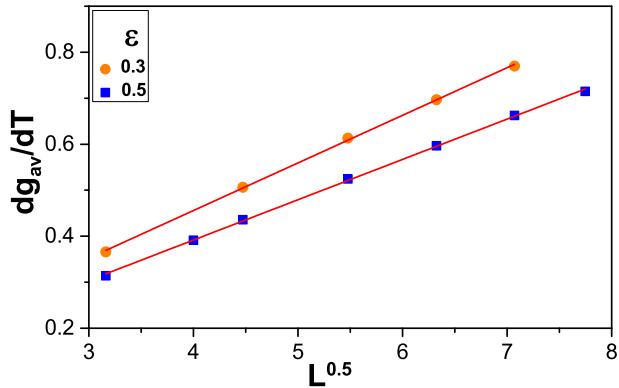}
\caption{(Color online) Slope $dg_{\rm av}/dT$ of the Binder cumulant vs.\ $L^{0.5}$ at the critical
temperature for $\epsilon=0.3$ and 0.5. The solid lines are linear fits.
          }
\label{fig:dgdTL}
\end{figure}

Finally, we analyze the slope $dg_{\rm av}/dT$ of the Binder cumulant at criticality.
It is expected to vary with system size as $dg_{\rm av}/dT \sim L^{1/\nu}$ with $\nu=2$. As is shown in Fig.\
\ref{fig:dgdTL}, our slopes indeed follow the power-law dependence predicted by the
strong-disorder renormalization group for both coupling strengths,
$\epsilon=0.3$ and 0.5.

\subsection{Strong coupling regime}
\label{subsec:MC_strong}

In the strong-coupling regime $\epsilon > \epsilon_c \approx 1.281$,
we perform simulations for coupling strengths $\epsilon=1.7$, 2.5, 3.5, and 5.
These simulations greatly suffer from
the fact that the embedded Wolff and Swendsen-Wang cluster algorithms are not valid for $\epsilon > 0.5$.
We are thus forced to employ the Metropolis single-spin algorithm. In this algorithm, the required
equilibration and measurement times increase significantly with system size, reaching several
hundred thousand sweeps for moderately large lattices. This severely limits the available sizes
and the accuracy of the results. For comparison, we also perform Wang-Landau simulations
but the available system sizes are restricted as well.

\begin{figure}
\includegraphics[width=8.2cm]{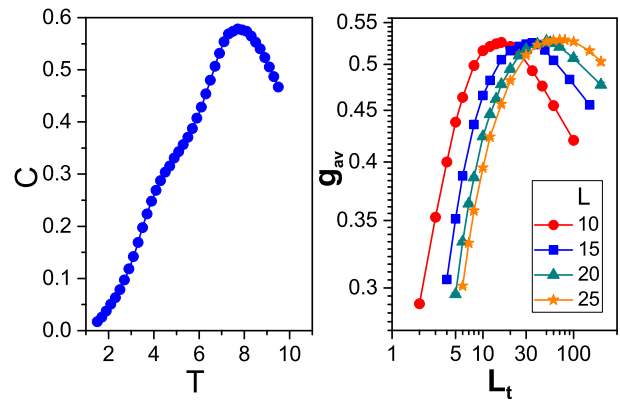}
\caption{(Color online) Left: Specific heat $C$ vs.\ classical temperature $T$ for
          $\epsilon=3.5$, system sizes $L=10$, $L_t=100$ and 5000 disorder configurations
          (using 140,000 Monte Carlo sweeps).  Notice two distinct
          peaks corresponding to two separate phase transitions. Right: Binder
          cumulant $g_{\rm av}$ as a function of $L_t$ for several $L$
         at the critical temperature $T_c=3.65$ for $\epsilon=1.7$. }
\label{fig:CT}
\end{figure}

As the classical Hamiltonian (\ref{eq:Hcl}) is not self-dual, we can expect a product phase
to appear for $\epsilon>\epsilon_c$. Indeed, for all studied $\epsilon$ values, we find
two distinct phase transitions. (This can already be seen from the specific heat data
shown in the left panel of Fig.\ \ref{fig:CT}.)
The product order parameter $p$, Eq.\ (\ref{eq:prod}), develops
at a higher temperature $T_c^p$ while the magnetization becomes nonzero only below a
lower temperature $T_c^m$ (see phase diagram in Fig.\ \ref{fig:pd}). In the following,
we look at these two transitions separately.

To analyze the transition between the product and Baxter phases (at which the magnetization
becomes critical), we use the same procedure based on the Binder cumulant $g_{\rm av}$
as in Sec.\ \ref{subsec:MC_weak}. The right panel of Fig.\ \ref{fig:CT} shows the Binder cumulant
at the estimated critical temperature $T_c^m=3.65$ for $\epsilon=1.7$ as a function of $L_t$ for several
$L$ between 10 and 25. As expected at criticality, the maximum value for each of the curves
does not depend on $L$. The figure also shows that the domes become broader with increasing
$L$, indicating non-power-law scaling. The largest spatial system size, $L=25$ requires
an enormous numerical effort, we averaged over 20,000 disorder configurations each using
700,000 Monte Carlo sweeps. Nonetheless the Binder cumulant at the right end of the dome
($L_t=200$) is not fully equilibrated as its value shifts with increasing number of sweeps.
Because of the limited system size range and the equilibration problems for the larger sizes
we are not able to quantitatively analyze the critical behavior of this transition.

\begin{figure}
\includegraphics[width=8.2cm]{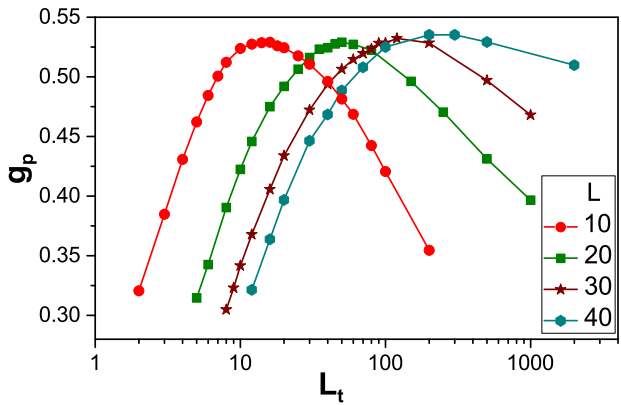}
\caption{(Color online) Product Binder cumulant $g_{p}$ as a function of $L_t$ for several $L$
        at the critical temperature $T_c^p=7.55$ for $\epsilon=3.5$.}
\label{fig:gLLtprod}
\end{figure}

\begin{figure}
\includegraphics[width=8.2cm]{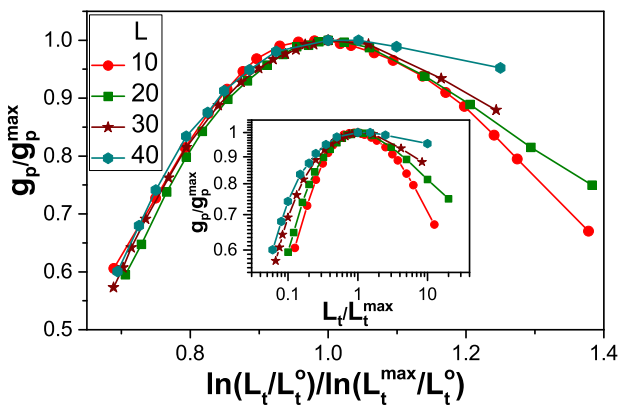}
\caption{(Color online) Scaling plot of the product Binder cumulant $g_p$ at $T_c^p=7.55$ for $\epsilon=3.5$. The symbols are
          the same as in Fig.\ \ref{fig:gLLtprod}
          Main panel: Activated scaling $g_{p}/g_{p}^{\rm max}$ vs.\  $\ln(L_t/L_t^0)/\ln(L_t^{\rm max}/L_t^0)$
          with $L_t^0=0.02$. Inset: Power-law scaling
          $g_{p}/g_{p}^{\rm max}$ vs.\ $L_t/L_t^{\rm max}$.
          }
\label{fig:gLLtprodscaling}
\end{figure}

Similar problems, though slightly less severe, also plague the transition between the
paramagnetic and product phases at which the product order parameter $p$ becomes critical.
Figure \ref{fig:gLLtprod} shows the Binder cumulant $g_p$ for the product order parameter
at the estimated critical temperature $T_c^p=7.55$ and $\epsilon=3.5$ as a function of $L_t$.
The maxima of the different curves are again independent of $L$, as expected at the critical
temperature. Moreover, the domes broaden with increasing system size. A scaling analysis of these
data is presented in Fig.\ \ref{fig:gLLtprodscaling}.
The inset shows that the behavior of $g_p$ is not compatible with conventional power-law scaling.
In contrast, the data scale reasonably well  when plotted versus $\ln(L_t/L_t^0)/\ln(L_t^{\rm max}/L_t^0)$
as shown in the main panel of the figure. This behavior is in agreement with activated
scaling in analogy to Eq.\ (\ref{eq:Binderscaling}) for the Binder cumulant $g_{\rm av}$ of the
magnetization. The deviations from data collapse for large $L_t$ (ie., at the right side of the domes)
stem from the fact that these systems do not equilibrate properly despite us using up to
500,000 Monte Carlo sweeps for each of the 20,000 disorder configurations (the $g_p$ values
still drift with increasing number of sweeps). This also prevents us from studying larger
system sizes.

If we ignore the small system size range and the equilibration problems and analyze (along the lines of Sec.\ \ref{subsec:MC_weak})
the system size dependencies of $L_t^{\rm max}$, the product order
parameter $p$, and its susceptibility $\chi_p$,  we obtain critical exponents that are roughly compatible with the
random transverse-field Ising universality class (as expected from the strong-disorder renormalization group).
We do not believe, however, that this constitutes a quantitative confirmation, and we
cannot rule out a different universality class with somewhat different critical exponents.

\section{Conclusions}
\label{sec:Conclusions}

In summary, we have studied the fate of the first-order quantum phase transition
in the three-color quantum Ashkin-Teller spin chain under the influence of quenched disorder.
To this end, we have mapped the random quantum Ashkin-Teller Hamiltonian onto a $(1+1)$-dimensional
classical Ashkin-Teller model with columnar disorder. We have then performed large-scale Monte Carlo
simulations for systems with up to 3.6 million lattice sites (10.8 million spins).
In agreement with the quantum version of the Aizenman-Wehr theorem, we have found that the
first-order transition of the clean system is rounded to a continuous one
in the presence of bond randomness.

For weak inter-color coupling $\epsilon$, efficient cluster Monte Carlo algorithms have
allowed us to simulate large systems. Our data for the quantum phase transition are in full
agreement with the results of the strong-disorder renormalization group calculation
\cite{GoswamiSchwabChakravarty08} that predicts universal critical behavior in the random
transverse-field Ising universality class. Specifically, we have confirmed for two different
values of $\epsilon$ the activated dynamical
scaling with a tunneling exponent $\psi=1/2$, the correlation length exponent $\nu=2$, and
the order parameter exponent $\beta=2-\phi$ with $\phi$ the golden mean. We have also
confirmed the behavior of the magnetic susceptibility.

In contrast, our simulations for large inter-color coupling $\epsilon$ have been restricted
to smaller system sizes, and they have suffered from equilibration problems because efficient
cluster algorithms are not available. Consequently, we have not been able to fully test the
renormalization group calculations in this regime. Our numerical data provide evidence
for activated dynamical scaling at the quantum phase transitions between the paramagnetic and
product phases as well as between the product and Baxter phases. For the latter transition we have
also determined rough estimates of the critical exponents and found them compatible with
the random transverse-field Ising universality class. However, a quantitative verification of
the critical behavior is beyond our current numerical capabilities.

Let us compare our results with earlier simulations. While our critical behavior (in the weak-coupling regime)
fully agrees with the random transverse-field Ising universality class, some exponents
calculated in Ref.\ \onlinecite{BellafardChakravarty16} show sizable deviations. This is
particularly interesting because the \emph{spatial} system sizes $L$ used in both simulations
are comparable (the largest $L$ in Ref.\ \onlinecite{BellafardChakravarty16} is actually
larger than ours). We believe that the results of Ref.\ \onlinecite{BellafardChakravarty16}
do not agree with the renormalization group predictions because the simulations are still
crossing over from the clean first-order transition to the disordered critical point,
probably because the chosen parameters lead to relatively weak disorder. This would mean
that the measured exponent values are effective rather than true asymptotic exponents.
Support for this hypothesis can be obtained from
comparing the dynamical scaling in the present paper and in Ref.\ \onlinecite{BellafardChakravarty16}. An infinite-randomness critical point features activated
dynamical scaling, i.e., the temporal system size $L_t$ scales exponentially with the spatial
size $L$ via $\ln(L_t) \sim L^\psi$. This implies that the conventional dynamical exponent $z=\infty$.
The optimal temporal system size (defined, e.g., via
the maximum of the Binder cumulant) therefore must increase very rapidly with $L$. Indeed, the
inset of Fig.\ \ref{fig:LLtmax} shows that $L_t^{\rm max}$ increases from 18 to about 2000
while $L$ varies only from 10 to 60.  The corresponding effective (scale-dependent) dynamical exponent
$z_{\rm eff}= d\ln(L_t^{\rm max})/d\ln(L)$ reaches almost 4 for the largest sizes.
In contrast, $L_t^{\rm max}$ reaches only 224 for $L=96$ in Ref.\ \onlinecite{BellafardChakravarty16}
and $z_{\rm eff}$ stays below 2, placing the system further away from the asymptotic
regime $z_{\rm eff} \to \infty$.

To conclude, as our numerical results (in the weak-coupling regime) fully agree with
the renormalization group predictions, we have not found any indications that the
\emph{asymptotic} critical behavior of the disordered system ``remembers'' the
first-order origin of the transition. This supports the expectation that the general classification
of disordered critical points developed in Refs.\
\onlinecite{VojtaSchmalian05,Vojta06,VojtaHoyos14} also holds for critical points emerging
from the rounding of first-order (quantum) phase transitions. However, the crossover from the clean to the
disordered behavior is certainly affected by the first-order nature of the clean transition.
The breakup length
beyond which phase coexistence is destroyed by domain formation increases with decreasing
disorder and may exceed the system size. For sufficiently weak disorder, the true asymptotic
behavior is then unobservable in both simulations and experiment.
This crossover will be even slower in $(2+1)$-dimensional systems because $d=2$
is the marginal dimension for the Aizenman-Wehr theorem, suggesting an exponential
dependence of the breakup length on the disorder strength.\cite{Binder83,GrinsteinMa83}

\section*{Acknowledgements}
This work was supported by the NSF under Grants No.\ DMR-1205803
and No. DMR-1506152. We thank Q. Zhu for sharing his Wang-Landau code.

\bibliographystyle{apsrev4-1}
\bibliography{../00bibtex/rareregions}

\begin{thebibliography}{65}%
\makeatletter
\providecommand \@ifxundefined [1]{%
 \@ifx{#1\undefined}
}%
\providecommand \@ifnum [1]{%
 \ifnum #1\expandafter \@firstoftwo
 \else \expandafter \@secondoftwo
 \fi
}%
\providecommand \@ifx [1]{%
 \ifx #1\expandafter \@firstoftwo
 \else \expandafter \@secondoftwo
 \fi
}%
\providecommand \natexlab [1]{#1}%
\providecommand \enquote  [1]{``#1''}%
\providecommand \bibnamefont  [1]{#1}%
\providecommand \bibfnamefont [1]{#1}%
\providecommand \citenamefont [1]{#1}%
\providecommand \href@noop [0]{\@secondoftwo}%
\providecommand \href [0]{\begingroup \@sanitize@url \@href}%
\providecommand \@href[1]{\@@startlink{#1}\@@href}%
\providecommand \@@href[1]{\endgroup#1\@@endlink}%
\providecommand \@sanitize@url [0]{\catcode `\\12\catcode `\$12\catcode
  `\&12\catcode `\#12\catcode `\^12\catcode `\_12\catcode `\%12\relax}%
\providecommand \@@startlink[1]{}%
\providecommand \@@endlink[0]{}%
\providecommand \url  [0]{\begingroup\@sanitize@url \@url }%
\providecommand \@url [1]{\endgroup\@href {#1}{\urlprefix }}%
\providecommand \urlprefix  [0]{URL }%
\providecommand \Eprint [0]{\href }%
\providecommand \doibase [0]{http://dx.doi.org/}%
\providecommand \selectlanguage [0]{\@gobble}%
\providecommand \bibinfo  [0]{\@secondoftwo}%
\providecommand \bibfield  [0]{\@secondoftwo}%
\providecommand \translation [1]{[#1]}%
\providecommand \BibitemOpen [0]{}%
\providecommand \bibitemStop [0]{}%
\providecommand \bibitemNoStop [0]{.\EOS\space}%
\providecommand \EOS [0]{\spacefactor3000\relax}%
\providecommand \BibitemShut  [1]{\csname bibitem#1\endcsname}%
\let\auto@bib@innerbib\@empty
\bibitem [{\citenamefont {Belitz}\ \emph {et~al.}(1997)\citenamefont {Belitz},
  \citenamefont {Kirkpatrick},\ and\ \citenamefont
  {Vojta}}]{BelitzKirkpatrickVojta97}%
  \BibitemOpen
  \bibfield  {author} {\bibinfo {author} {\bibfnamefont {D.}~\bibnamefont
  {Belitz}}, \bibinfo {author} {\bibfnamefont {T.~R.}\ \bibnamefont
  {Kirkpatrick}}, \ and\ \bibinfo {author} {\bibfnamefont {T.}~\bibnamefont
  {Vojta}},\ }\href@noop {} {\bibfield  {journal} {\bibinfo  {journal} {Phys.
  Rev. B}\ }\textbf {\bibinfo {volume} {55}},\ \bibinfo {pages} {9452}
  (\bibinfo {year} {1997})}\BibitemShut {NoStop}%
\bibitem [{\citenamefont {Belitz}\ \emph {et~al.}(1999)\citenamefont {Belitz},
  \citenamefont {Kirkpatrick},\ and\ \citenamefont
  {Vojta}}]{BelitzKirkpatrickVojta99}%
  \BibitemOpen
  \bibfield  {author} {\bibinfo {author} {\bibfnamefont {D.}~\bibnamefont
  {Belitz}}, \bibinfo {author} {\bibfnamefont {T.~R.}\ \bibnamefont
  {Kirkpatrick}}, \ and\ \bibinfo {author} {\bibfnamefont {T.}~\bibnamefont
  {Vojta}},\ }\href@noop {} {\bibfield  {journal} {\bibinfo  {journal} {Phys.
  Rev. Lett.}\ }\textbf {\bibinfo {volume} {82}},\ \bibinfo {pages} {4707}
  (\bibinfo {year} {1999})}\BibitemShut {NoStop}%
\bibitem [{\citenamefont {Belitz}\ \emph {et~al.}(2005)\citenamefont {Belitz},
  \citenamefont {Kirkpatrick},\ and\ \citenamefont
  {Vojta}}]{BelitzKirkpatrickVojta05}%
  \BibitemOpen
  \bibfield  {author} {\bibinfo {author} {\bibfnamefont {D.}~\bibnamefont
  {Belitz}}, \bibinfo {author} {\bibfnamefont {T.~R.}\ \bibnamefont
  {Kirkpatrick}}, \ and\ \bibinfo {author} {\bibfnamefont {T.}~\bibnamefont
  {Vojta}},\ }\href@noop {} {\bibfield  {journal} {\bibinfo  {journal} {Rev.
  Mod. Phys.}\ }\textbf {\bibinfo {volume} {77}},\ \bibinfo {pages} {579}
  (\bibinfo {year} {2005})}\BibitemShut {NoStop}%
\bibitem [{\citenamefont {Brando}\ \emph {et~al.}(2016)\citenamefont {Brando},
  \citenamefont {Belitz}, \citenamefont {Grosche},\ and\ \citenamefont
  {Kirkpatrick}}]{BrandoBelitzGroscheKirkpatrick16}%
  \BibitemOpen
  \bibfield  {author} {\bibinfo {author} {\bibfnamefont {M.}~\bibnamefont
  {Brando}}, \bibinfo {author} {\bibfnamefont {D.}~\bibnamefont {Belitz}},
  \bibinfo {author} {\bibfnamefont {F.~M.}\ \bibnamefont {Grosche}}, \ and\
  \bibinfo {author} {\bibfnamefont {T.~R.}\ \bibnamefont {Kirkpatrick}},\
  }\href {\doibase 10.1103/RevModPhys.88.025006} {\bibfield  {journal}
  {\bibinfo  {journal} {Rev. Mod. Phys.}\ }\textbf {\bibinfo {volume} {88}},\
  \bibinfo {pages} {025006} (\bibinfo {year} {2016})}\BibitemShut {NoStop}%
\bibitem [{\citenamefont {Fisher}(1992)}]{Fisher92}%
  \BibitemOpen
  \bibfield  {author} {\bibinfo {author} {\bibfnamefont {D.~S.}\ \bibnamefont
  {Fisher}},\ }\href@noop {} {\bibfield  {journal} {\bibinfo  {journal} {Phys.
  Rev. Lett.}\ }\textbf {\bibinfo {volume} {69}},\ \bibinfo {pages} {534}
  (\bibinfo {year} {1992})}\BibitemShut {NoStop}%
\bibitem [{\citenamefont {Fisher}(1995)}]{Fisher95}%
  \BibitemOpen
  \bibfield  {author} {\bibinfo {author} {\bibfnamefont {D.~S.}\ \bibnamefont
  {Fisher}},\ }\href {\doibase 10.1103/PhysRevB.51.6411} {\bibfield  {journal}
  {\bibinfo  {journal} {Phys. Rev. B}\ }\textbf {\bibinfo {volume} {51}},\
  \bibinfo {pages} {6411} (\bibinfo {year} {1995})}\BibitemShut {NoStop}%
\bibitem [{\citenamefont {Motrunich}\ \emph {et~al.}(2000)\citenamefont
  {Motrunich}, \citenamefont {Mau}, \citenamefont {Huse},\ and\ \citenamefont
  {Fisher}}]{MMHF00}%
  \BibitemOpen
  \bibfield  {author} {\bibinfo {author} {\bibfnamefont {O.}~\bibnamefont
  {Motrunich}}, \bibinfo {author} {\bibfnamefont {S.~C.}\ \bibnamefont {Mau}},
  \bibinfo {author} {\bibfnamefont {D.~A.}\ \bibnamefont {Huse}}, \ and\
  \bibinfo {author} {\bibfnamefont {D.~S.}\ \bibnamefont {Fisher}},\ }\href
  {\doibase 10.1103/PhysRevB.61.1160} {\bibfield  {journal} {\bibinfo
  {journal} {Phys. Rev. B}\ }\textbf {\bibinfo {volume} {61}},\ \bibinfo
  {pages} {1160} (\bibinfo {year} {2000})}\BibitemShut {NoStop}%
\bibitem [{\citenamefont {Hoyos}\ \emph {et~al.}(2007)\citenamefont {Hoyos},
  \citenamefont {Kotabage},\ and\ \citenamefont
  {Vojta}}]{HoyosKotabageVojta07}%
  \BibitemOpen
  \bibfield  {author} {\bibinfo {author} {\bibfnamefont {J.~A.}\ \bibnamefont
  {Hoyos}}, \bibinfo {author} {\bibfnamefont {C.}~\bibnamefont {Kotabage}}, \
  and\ \bibinfo {author} {\bibfnamefont {T.}~\bibnamefont {Vojta}},\ }\href
  {\doibase 10.1103/PhysRevLett.99.230601} {\bibfield  {journal} {\bibinfo
  {journal} {Phys. Rev. Lett.}\ }\textbf {\bibinfo {volume} {99}},\ \bibinfo
  {pages} {230601} (\bibinfo {year} {2007})}\BibitemShut {NoStop}%
\bibitem [{\citenamefont {Vojta}\ \emph {et~al.}(2009)\citenamefont {Vojta},
  \citenamefont {Kotabage},\ and\ \citenamefont
  {Hoyos}}]{VojtaKotabageHoyos09}%
  \BibitemOpen
  \bibfield  {author} {\bibinfo {author} {\bibfnamefont {T.}~\bibnamefont
  {Vojta}}, \bibinfo {author} {\bibfnamefont {C.}~\bibnamefont {Kotabage}}, \
  and\ \bibinfo {author} {\bibfnamefont {J.~A.}\ \bibnamefont {Hoyos}},\ }\href
  {\doibase 10.1103/PhysRevB.79.024401} {\bibfield  {journal} {\bibinfo
  {journal} {Phys. Rev. B}\ }\textbf {\bibinfo {volume} {79}},\ \bibinfo
  {pages} {024401} (\bibinfo {year} {2009})}\BibitemShut {NoStop}%
\bibitem [{\citenamefont {Thill}\ and\ \citenamefont
  {Huse}(1995)}]{ThillHuse95}%
  \BibitemOpen
  \bibfield  {author} {\bibinfo {author} {\bibfnamefont {M.}~\bibnamefont
  {Thill}}\ and\ \bibinfo {author} {\bibfnamefont {D.~A.}\ \bibnamefont
  {Huse}},\ }\href {\doibase 10.1016/0378-4371(94)00247-Q} {\bibfield
  {journal} {\bibinfo  {journal} {Physica A}\ }\textbf {\bibinfo {volume}
  {214}},\ \bibinfo {pages} {321} (\bibinfo {year} {1995})}\BibitemShut
  {NoStop}%
\bibitem [{\citenamefont {Rieger}\ and\ \citenamefont
  {Young}(1996)}]{RiegerYoung96}%
  \BibitemOpen
  \bibfield  {author} {\bibinfo {author} {\bibfnamefont {H.}~\bibnamefont
  {Rieger}}\ and\ \bibinfo {author} {\bibfnamefont {A.~P.}\ \bibnamefont
  {Young}},\ }\href {\doibase 10.1103/PhysRevB.54.3328} {\bibfield  {journal}
  {\bibinfo  {journal} {Phys. Rev. B}\ }\textbf {\bibinfo {volume} {54}},\
  \bibinfo {pages} {3328} (\bibinfo {year} {1996})}\BibitemShut {NoStop}%
\bibitem [{\citenamefont {Young}\ and\ \citenamefont
  {Rieger}(1996)}]{YoungRieger96}%
  \BibitemOpen
  \bibfield  {author} {\bibinfo {author} {\bibfnamefont {A.~P.}\ \bibnamefont
  {Young}}\ and\ \bibinfo {author} {\bibfnamefont {H.}~\bibnamefont {Rieger}},\
  }\href@noop {} {\bibfield  {journal} {\bibinfo  {journal} {Phys. Rev. B}\
  }\textbf {\bibinfo {volume} {53}},\ \bibinfo {pages} {8486} (\bibinfo {year}
  {1996})}\BibitemShut {NoStop}%
\bibitem [{\citenamefont {Vojta}(2003)}]{Vojta03a}%
  \BibitemOpen
  \bibfield  {author} {\bibinfo {author} {\bibfnamefont {T.}~\bibnamefont
  {Vojta}},\ }\href@noop {} {\bibfield  {journal} {\bibinfo  {journal} {Phys.
  Rev. Lett.}\ }\textbf {\bibinfo {volume} {90}},\ \bibinfo {pages} {107202}
  (\bibinfo {year} {2003})}\BibitemShut {NoStop}%
\bibitem [{\citenamefont {Hoyos}\ and\ \citenamefont
  {Vojta}(2008)}]{HoyosVojta08}%
  \BibitemOpen
  \bibfield  {author} {\bibinfo {author} {\bibfnamefont {J.~A.}\ \bibnamefont
  {Hoyos}}\ and\ \bibinfo {author} {\bibfnamefont {T.}~\bibnamefont {Vojta}},\
  }\href@noop {} {\bibfield  {journal} {\bibinfo  {journal} {Phys. Rev. Lett.}\
  }\textbf {\bibinfo {volume} {100}},\ \bibinfo {pages} {240601} (\bibinfo
  {year} {2008})}\BibitemShut {NoStop}%
\bibitem [{\citenamefont {Guo}\ \emph {et~al.}(2007)\citenamefont {Guo},
  \citenamefont {Young}, \citenamefont {Macaluso}, \citenamefont {Browne},
  \citenamefont {Henderson}, \citenamefont {Chan}, \citenamefont {Henry},\ and\
  \citenamefont {DiTusa}}]{GYMBHCHD07}%
  \BibitemOpen
  \bibfield  {author} {\bibinfo {author} {\bibfnamefont {S.}~\bibnamefont
  {Guo}}, \bibinfo {author} {\bibfnamefont {D.~P.}\ \bibnamefont {Young}},
  \bibinfo {author} {\bibfnamefont {R.~T.}\ \bibnamefont {Macaluso}}, \bibinfo
  {author} {\bibfnamefont {D.~A.}\ \bibnamefont {Browne}}, \bibinfo {author}
  {\bibfnamefont {N.~L.}\ \bibnamefont {Henderson}}, \bibinfo {author}
  {\bibfnamefont {J.~Y.}\ \bibnamefont {Chan}}, \bibinfo {author}
  {\bibfnamefont {L.}~\bibnamefont {Henry}}, \ and\ \bibinfo {author}
  {\bibfnamefont {J.~F.}\ \bibnamefont {DiTusa}},\ }\href@noop {} {\bibfield
  {journal} {\bibinfo  {journal} {Phys. Rev. Lett.}\ }\textbf {\bibinfo
  {volume} {100}},\ \bibinfo {pages} {017209} (\bibinfo {year}
  {2007})}\BibitemShut {NoStop}%
\bibitem [{\citenamefont {Guo}\ \emph {et~al.}(2010)\citenamefont {Guo},
  \citenamefont {Young}, \citenamefont {Macaluso}, \citenamefont {Browne},
  \citenamefont {Henderson}, \citenamefont {Chan}, \citenamefont {Henry},\ and\
  \citenamefont {DiTusa}}]{GYMBHCHD10a}%
  \BibitemOpen
  \bibfield  {author} {\bibinfo {author} {\bibfnamefont {S.}~\bibnamefont
  {Guo}}, \bibinfo {author} {\bibfnamefont {D.~P.}\ \bibnamefont {Young}},
  \bibinfo {author} {\bibfnamefont {R.~T.}\ \bibnamefont {Macaluso}}, \bibinfo
  {author} {\bibfnamefont {D.~A.}\ \bibnamefont {Browne}}, \bibinfo {author}
  {\bibfnamefont {N.~L.}\ \bibnamefont {Henderson}}, \bibinfo {author}
  {\bibfnamefont {J.~Y.}\ \bibnamefont {Chan}}, \bibinfo {author}
  {\bibfnamefont {L.~L.}\ \bibnamefont {Henry}}, \ and\ \bibinfo {author}
  {\bibfnamefont {J.~F.}\ \bibnamefont {DiTusa}},\ }\href@noop {} {\bibfield
  {journal} {\bibinfo  {journal} {Phys. Rev. B}\ }\textbf {\bibinfo {volume}
  {81}},\ \bibinfo {pages} {144423} (\bibinfo {year} {2010})}\BibitemShut
  {NoStop}%
\bibitem [{\citenamefont {Westerkamp}\ \emph {et~al.}(2009)\citenamefont
  {Westerkamp}, \citenamefont {Deppe}, \citenamefont {K{\"u}chler},
  \citenamefont {Brando}, \citenamefont {Geibel}, \citenamefont {Gegenwart},
  \citenamefont {Pikul},\ and\ \citenamefont {Steglich}}]{Westerkampetal09}%
  \BibitemOpen
  \bibfield  {author} {\bibinfo {author} {\bibfnamefont {T.}~\bibnamefont
  {Westerkamp}}, \bibinfo {author} {\bibfnamefont {M.}~\bibnamefont {Deppe}},
  \bibinfo {author} {\bibfnamefont {R.}~\bibnamefont {K{\"u}chler}}, \bibinfo
  {author} {\bibfnamefont {M.}~\bibnamefont {Brando}}, \bibinfo {author}
  {\bibfnamefont {C.}~\bibnamefont {Geibel}}, \bibinfo {author} {\bibfnamefont
  {P.}~\bibnamefont {Gegenwart}}, \bibinfo {author} {\bibfnamefont {A.~P.}\
  \bibnamefont {Pikul}}, \ and\ \bibinfo {author} {\bibfnamefont
  {F.}~\bibnamefont {Steglich}},\ }\href@noop {} {\bibfield  {journal}
  {\bibinfo  {journal} {Phys. Rev. Lett.}\ }\textbf {\bibinfo {volume} {102}},\
  \bibinfo {pages} {206404} (\bibinfo {year} {2009})}\BibitemShut {NoStop}%
\bibitem [{\citenamefont {Ubaid-Kassis}\ \emph {et~al.}(2010)\citenamefont
  {Ubaid-Kassis}, \citenamefont {Vojta},\ and\ \citenamefont
  {Schroeder}}]{UbaidKassisVojtaSchroeder10}%
  \BibitemOpen
  \bibfield  {author} {\bibinfo {author} {\bibfnamefont {S.}~\bibnamefont
  {Ubaid-Kassis}}, \bibinfo {author} {\bibfnamefont {T.}~\bibnamefont {Vojta}},
  \ and\ \bibinfo {author} {\bibfnamefont {A.}~\bibnamefont {Schroeder}},\
  }\href@noop {} {\bibfield  {journal} {\bibinfo  {journal} {Phys. Rev. Lett.}\
  }\textbf {\bibinfo {volume} {104}},\ \bibinfo {pages} {066402} (\bibinfo
  {year} {2010})}\BibitemShut {NoStop}%
\bibitem [{\citenamefont {Demk\'o}\ \emph {et~al.}(2012)\citenamefont
  {Demk\'o}, \citenamefont {Bord\'acs}, \citenamefont {Vojta}, \citenamefont
  {Nozadze}, \citenamefont {Hrahsheh}, \citenamefont {Svoboda}, \citenamefont
  {D\'ora}, \citenamefont {Yamada}, \citenamefont {Kawasaki}, \citenamefont
  {Tokura},\ and\ \citenamefont {K\'ezsm\'arki}}]{Demkoetal12}%
  \BibitemOpen
  \bibfield  {author} {\bibinfo {author} {\bibfnamefont {L.}~\bibnamefont
  {Demk\'o}}, \bibinfo {author} {\bibfnamefont {S.}~\bibnamefont {Bord\'acs}},
  \bibinfo {author} {\bibfnamefont {T.}~\bibnamefont {Vojta}}, \bibinfo
  {author} {\bibfnamefont {D.}~\bibnamefont {Nozadze}}, \bibinfo {author}
  {\bibfnamefont {F.}~\bibnamefont {Hrahsheh}}, \bibinfo {author}
  {\bibfnamefont {C.}~\bibnamefont {Svoboda}}, \bibinfo {author} {\bibfnamefont
  {B.}~\bibnamefont {D\'ora}}, \bibinfo {author} {\bibfnamefont
  {H.}~\bibnamefont {Yamada}}, \bibinfo {author} {\bibfnamefont
  {M.}~\bibnamefont {Kawasaki}}, \bibinfo {author} {\bibfnamefont
  {Y.}~\bibnamefont {Tokura}}, \ and\ \bibinfo {author} {\bibfnamefont
  {I.}~\bibnamefont {K\'ezsm\'arki}},\ }\href {\doibase
  10.1103/PhysRevLett.108.185701} {\bibfield  {journal} {\bibinfo  {journal}
  {Phys. Rev. Lett.}\ }\textbf {\bibinfo {volume} {108}},\ \bibinfo {pages}
  {185701} (\bibinfo {year} {2012})}\BibitemShut {NoStop}%
\bibitem [{\citenamefont {Vojta}\ and\ \citenamefont
  {Schmalian}(2005)}]{VojtaSchmalian05}%
  \BibitemOpen
  \bibfield  {author} {\bibinfo {author} {\bibfnamefont {T.}~\bibnamefont
  {Vojta}}\ and\ \bibinfo {author} {\bibfnamefont {J.}~\bibnamefont
  {Schmalian}},\ }\href {\doibase 10.1103/PhysRevB.72.045438} {\bibfield
  {journal} {\bibinfo  {journal} {Phys. Rev. B}\ }\textbf {\bibinfo {volume}
  {72}},\ \bibinfo {pages} {045438} (\bibinfo {year} {2005})}\BibitemShut
  {NoStop}%
\bibitem [{\citenamefont {Vojta}\ and\ \citenamefont
  {Hoyos}(2014)}]{VojtaHoyos14}%
  \BibitemOpen
  \bibfield  {author} {\bibinfo {author} {\bibfnamefont {T.}~\bibnamefont
  {Vojta}}\ and\ \bibinfo {author} {\bibfnamefont {J.~A.}\ \bibnamefont
  {Hoyos}},\ }\href {\doibase 10.1103/PhysRevLett.112.075702} {\bibfield
  {journal} {\bibinfo  {journal} {Phys. Rev. Lett.}\ }\textbf {\bibinfo
  {volume} {112}},\ \bibinfo {pages} {075702} (\bibinfo {year}
  {2014})}\BibitemShut {NoStop}%
\bibitem [{\citenamefont {Vojta}(2006)}]{Vojta06}%
  \BibitemOpen
  \bibfield  {author} {\bibinfo {author} {\bibfnamefont {T.}~\bibnamefont
  {Vojta}},\ }\href {\doibase 10.1088/0305-4470/39/22/R01} {\bibfield
  {journal} {\bibinfo  {journal} {J. Phys. A}\ }\textbf {\bibinfo {volume}
  {39}},\ \bibinfo {pages} {R143} (\bibinfo {year} {2006})}\BibitemShut
  {NoStop}%
\bibitem [{\citenamefont {Vojta}(2010)}]{Vojta10}%
  \BibitemOpen
  \bibfield  {author} {\bibinfo {author} {\bibfnamefont {T.}~\bibnamefont
  {Vojta}},\ }\href {\doibase 10.1007/s10909-010-0205-4} {\bibfield  {journal}
  {\bibinfo  {journal} {J. Low Temp. Phys.}\ }\textbf {\bibinfo {volume}
  {161}},\ \bibinfo {pages} {299} (\bibinfo {year} {2010})}\BibitemShut
  {NoStop}%
\bibitem [{\citenamefont {Vojta}(2014)}]{Vojta14}%
  \BibitemOpen
  \bibfield  {author} {\bibinfo {author} {\bibfnamefont {T.}~\bibnamefont
  {Vojta}},\ }\href {http://stacks.iop.org/1742-6596/529/i=1/a=012016}
  {\bibfield  {journal} {\bibinfo  {journal} {J. Phys. Conf. Series}\ }\textbf
  {\bibinfo {volume} {529}},\ \bibinfo {pages} {012016} (\bibinfo {year}
  {2014})}\BibitemShut {NoStop}%
\bibitem [{\citenamefont {Greenblatt}\ \emph {et~al.}(2009)\citenamefont
  {Greenblatt}, \citenamefont {Aizenman},\ and\ \citenamefont
  {Lebowitz}}]{GreenblattAizenmanLebowitz09}%
  \BibitemOpen
  \bibfield  {author} {\bibinfo {author} {\bibfnamefont {R.~L.}\ \bibnamefont
  {Greenblatt}}, \bibinfo {author} {\bibfnamefont {M.}~\bibnamefont
  {Aizenman}}, \ and\ \bibinfo {author} {\bibfnamefont {J.~L.}\ \bibnamefont
  {Lebowitz}},\ }\href {\doibase 10.1103/PhysRevLett.103.197201} {\bibfield
  {journal} {\bibinfo  {journal} {Phys. Rev. Lett.}\ }\textbf {\bibinfo
  {volume} {103}},\ \bibinfo {pages} {197201} (\bibinfo {year}
  {2009})}\BibitemShut {NoStop}%
\bibitem [{\citenamefont {Aizenman}\ \emph {et~al.}(2012)\citenamefont
  {Aizenman}, \citenamefont {Greenblatt},\ and\ \citenamefont
  {Lebowitz}}]{AizenmanGreenblattLebowitz12}%
  \BibitemOpen
  \bibfield  {author} {\bibinfo {author} {\bibfnamefont {M.}~\bibnamefont
  {Aizenman}}, \bibinfo {author} {\bibfnamefont {R.~L.}\ \bibnamefont
  {Greenblatt}}, \ and\ \bibinfo {author} {\bibfnamefont {J.~L.}\ \bibnamefont
  {Lebowitz}},\ }\href {\doibase http://dx.doi.org/10.1063/1.3679069}
  {\bibfield  {journal} {\bibinfo  {journal} {J. Math. Phys.}\ }\textbf
  {\bibinfo {volume} {53}},\ \bibinfo {eid} {023301} (\bibinfo {year}
  {2012})}\BibitemShut {NoStop}%
\bibitem [{\citenamefont {Imry}\ and\ \citenamefont
  {Wortis}(1979)}]{ImryWortis79}%
  \BibitemOpen
  \bibfield  {author} {\bibinfo {author} {\bibfnamefont {Y.}~\bibnamefont
  {Imry}}\ and\ \bibinfo {author} {\bibfnamefont {M.}~\bibnamefont {Wortis}},\
  }\href {\doibase 10.1103/PhysRevB.19.3580} {\bibfield  {journal} {\bibinfo
  {journal} {Phys. Rev. B}\ }\textbf {\bibinfo {volume} {19}},\ \bibinfo
  {pages} {3580} (\bibinfo {year} {1979})}\BibitemShut {NoStop}%
\bibitem [{\citenamefont {Hui}\ and\ \citenamefont
  {Berker}(1989)}]{HuiBerker89}%
  \BibitemOpen
  \bibfield  {author} {\bibinfo {author} {\bibfnamefont {K.}~\bibnamefont
  {Hui}}\ and\ \bibinfo {author} {\bibfnamefont {A.~N.}\ \bibnamefont
  {Berker}},\ }\href {\doibase 10.1103/PhysRevLett.62.2507} {\bibfield
  {journal} {\bibinfo  {journal} {Phys. Rev. Lett.}\ }\textbf {\bibinfo
  {volume} {62}},\ \bibinfo {pages} {2507} (\bibinfo {year}
  {1989})}\BibitemShut {NoStop}%
\bibitem [{\citenamefont {Aizenman}\ and\ \citenamefont
  {Wehr}(1989)}]{AizenmanWehr89}%
  \BibitemOpen
  \bibfield  {author} {\bibinfo {author} {\bibfnamefont {M.}~\bibnamefont
  {Aizenman}}\ and\ \bibinfo {author} {\bibfnamefont {J.}~\bibnamefont
  {Wehr}},\ }\href {\doibase 10.1103/PhysRevLett.62.2503} {\bibfield  {journal}
  {\bibinfo  {journal} {Phys. Rev. Lett.}\ }\textbf {\bibinfo {volume} {62}},\
  \bibinfo {pages} {2503} (\bibinfo {year} {1989})}\BibitemShut {NoStop}%
\bibitem [{\citenamefont {Senthil}\ and\ \citenamefont
  {Majumdar}(1996)}]{SenthilMajumdar96}%
  \BibitemOpen
  \bibfield  {author} {\bibinfo {author} {\bibfnamefont {T.}~\bibnamefont
  {Senthil}}\ and\ \bibinfo {author} {\bibfnamefont {S.~N.}\ \bibnamefont
  {Majumdar}},\ }\href@noop {} {\bibfield  {journal} {\bibinfo  {journal}
  {Phys. Rev. Lett.}\ }\textbf {\bibinfo {volume} {76}},\ \bibinfo {pages}
  {3001} (\bibinfo {year} {1996})}\BibitemShut {NoStop}%
\bibitem [{\citenamefont {Goswami}\ \emph {et~al.}(2008)\citenamefont
  {Goswami}, \citenamefont {Schwab},\ and\ \citenamefont
  {Chakravarty}}]{GoswamiSchwabChakravarty08}%
  \BibitemOpen
  \bibfield  {author} {\bibinfo {author} {\bibfnamefont {P.}~\bibnamefont
  {Goswami}}, \bibinfo {author} {\bibfnamefont {D.}~\bibnamefont {Schwab}}, \
  and\ \bibinfo {author} {\bibfnamefont {S.}~\bibnamefont {Chakravarty}},\
  }\href@noop {} {\bibfield  {journal} {\bibinfo  {journal} {Phys. Rev. Lett.}\
  }\textbf {\bibinfo {volume} {100}},\ \bibinfo {pages} {015703} (\bibinfo
  {year} {2008})}\BibitemShut {NoStop}%
\bibitem [{\citenamefont {Fradkin}(1984)}]{Fradkin84}%
  \BibitemOpen
  \bibfield  {author} {\bibinfo {author} {\bibfnamefont {E.}~\bibnamefont
  {Fradkin}},\ }\href {\doibase 10.1103/PhysRevLett.53.1967} {\bibfield
  {journal} {\bibinfo  {journal} {Phys. Rev. Lett.}\ }\textbf {\bibinfo
  {volume} {53}},\ \bibinfo {pages} {1967} (\bibinfo {year}
  {1984})}\BibitemShut {NoStop}%
\bibitem [{\citenamefont {Shankar}(1985)}]{Shankar85}%
  \BibitemOpen
  \bibfield  {author} {\bibinfo {author} {\bibfnamefont {R.}~\bibnamefont
  {Shankar}},\ }\href {\doibase 10.1103/PhysRevLett.55.453} {\bibfield
  {journal} {\bibinfo  {journal} {Phys. Rev. Lett.}\ }\textbf {\bibinfo
  {volume} {55}},\ \bibinfo {pages} {453} (\bibinfo {year} {1985})}\BibitemShut
  {NoStop}%
\bibitem [{\citenamefont {Hrahsheh}\ \emph {et~al.}(2012)\citenamefont
  {Hrahsheh}, \citenamefont {Hoyos},\ and\ \citenamefont
  {Vojta}}]{HrahshehHoyosVojta12}%
  \BibitemOpen
  \bibfield  {author} {\bibinfo {author} {\bibfnamefont {F.}~\bibnamefont
  {Hrahsheh}}, \bibinfo {author} {\bibfnamefont {J.~A.}\ \bibnamefont {Hoyos}},
  \ and\ \bibinfo {author} {\bibfnamefont {T.}~\bibnamefont {Vojta}},\ }\href
  {\doibase 10.1103/PhysRevB.86.214204} {\bibfield  {journal} {\bibinfo
  {journal} {Phys. Rev. B}\ }\textbf {\bibinfo {volume} {86}},\ \bibinfo
  {pages} {214204} (\bibinfo {year} {2012})}\BibitemShut {NoStop}%
\bibitem [{\citenamefont {Barghathi}\ \emph {et~al.}(2015)\citenamefont
  {Barghathi}, \citenamefont {Hrahsheh}, \citenamefont {Hoyos}, \citenamefont
  {Narayanan},\ and\ \citenamefont {Vojta}}]{Barghathietal14}%
  \BibitemOpen
  \bibfield  {author} {\bibinfo {author} {\bibfnamefont {H.}~\bibnamefont
  {Barghathi}}, \bibinfo {author} {\bibfnamefont {F.}~\bibnamefont {Hrahsheh}},
  \bibinfo {author} {\bibfnamefont {J.~A.}\ \bibnamefont {Hoyos}}, \bibinfo
  {author} {\bibfnamefont {R.}~\bibnamefont {Narayanan}}, \ and\ \bibinfo
  {author} {\bibfnamefont {T.}~\bibnamefont {Vojta}},\ }\href@noop {}
  {\bibfield  {journal} {\bibinfo  {journal} {Phys. Scr.}\ }\textbf {\bibinfo
  {volume} {T165}},\ \bibinfo {pages} {014040} (\bibinfo {year}
  {2015})}\BibitemShut {NoStop}%
\bibitem [{\citenamefont {Igloi}\ and\ \citenamefont
  {Monthus}(2005)}]{IgloiMonthus05}%
  \BibitemOpen
  \bibfield  {author} {\bibinfo {author} {\bibfnamefont {F.}~\bibnamefont
  {Igloi}}\ and\ \bibinfo {author} {\bibfnamefont {C.}~\bibnamefont
  {Monthus}},\ }\href@noop {} {\bibfield  {journal} {\bibinfo  {journal} {Phys.
  Rep.}\ }\textbf {\bibinfo {volume} {412}},\ \bibinfo {pages} {277} (\bibinfo
  {year} {2005})}\BibitemShut {NoStop}%
\bibitem [{\citenamefont {Bellafard}\ and\ \citenamefont
  {Chakravarty}(2016)}]{BellafardChakravarty16}%
  \BibitemOpen
  \bibfield  {author} {\bibinfo {author} {\bibfnamefont {A.}~\bibnamefont
  {Bellafard}}\ and\ \bibinfo {author} {\bibfnamefont {S.}~\bibnamefont
  {Chakravarty}},\ }\href {\doibase 10.1103/PhysRevB.94.094408} {\bibfield
  {journal} {\bibinfo  {journal} {Phys. Rev. B}\ }\textbf {\bibinfo {volume}
  {94}},\ \bibinfo {pages} {094408} (\bibinfo {year} {2016})}\BibitemShut
  {NoStop}%
\bibitem [{\citenamefont {Grest}\ and\ \citenamefont
  {Widom}(1981)}]{GrestWidom81}%
  \BibitemOpen
  \bibfield  {author} {\bibinfo {author} {\bibfnamefont {G.~S.}\ \bibnamefont
  {Grest}}\ and\ \bibinfo {author} {\bibfnamefont {M.}~\bibnamefont {Widom}},\
  }\href {\doibase 10.1103/PhysRevB.24.6508} {\bibfield  {journal} {\bibinfo
  {journal} {Phys. Rev. B}\ }\textbf {\bibinfo {volume} {24}},\ \bibinfo
  {pages} {6508} (\bibinfo {year} {1981})}\BibitemShut {NoStop}%
\bibitem [{\citenamefont {Ashkin}\ and\ \citenamefont
  {Teller}(1943)}]{AshkinTeller43}%
  \BibitemOpen
  \bibfield  {author} {\bibinfo {author} {\bibfnamefont {J.}~\bibnamefont
  {Ashkin}}\ and\ \bibinfo {author} {\bibfnamefont {E.}~\bibnamefont
  {Teller}},\ }\href {\doibase 10.1103/PhysRev.64.178} {\bibfield  {journal}
  {\bibinfo  {journal} {Phys. Rev.}\ }\textbf {\bibinfo {volume} {64}},\
  \bibinfo {pages} {178} (\bibinfo {year} {1943})}\BibitemShut {NoStop}%
\bibitem [{\citenamefont {Bak}\ \emph {et~al.}(1985)\citenamefont {Bak},
  \citenamefont {Kleban}, \citenamefont {Unertl}, \citenamefont {Ochab},
  \citenamefont {Akinci}, \citenamefont {Bartelt},\ and\ \citenamefont
  {Einstein}}]{Baketal85}%
  \BibitemOpen
  \bibfield  {author} {\bibinfo {author} {\bibfnamefont {P.}~\bibnamefont
  {Bak}}, \bibinfo {author} {\bibfnamefont {P.}~\bibnamefont {Kleban}},
  \bibinfo {author} {\bibfnamefont {W.~N.}\ \bibnamefont {Unertl}}, \bibinfo
  {author} {\bibfnamefont {J.}~\bibnamefont {Ochab}}, \bibinfo {author}
  {\bibfnamefont {G.}~\bibnamefont {Akinci}}, \bibinfo {author} {\bibfnamefont
  {N.~C.}\ \bibnamefont {Bartelt}}, \ and\ \bibinfo {author} {\bibfnamefont
  {T.~L.}\ \bibnamefont {Einstein}},\ }\href {\doibase
  10.1103/PhysRevLett.54.1539} {\bibfield  {journal} {\bibinfo  {journal}
  {Phys. Rev. Lett.}\ }\textbf {\bibinfo {volume} {54}},\ \bibinfo {pages}
  {1539} (\bibinfo {year} {1985})}\BibitemShut {NoStop}%
\bibitem [{\citenamefont {Aji}\ and\ \citenamefont {Varma}(2007)}]{AjiVarma07}%
  \BibitemOpen
  \bibfield  {author} {\bibinfo {author} {\bibfnamefont {V.}~\bibnamefont
  {Aji}}\ and\ \bibinfo {author} {\bibfnamefont {C.~M.}\ \bibnamefont
  {Varma}},\ }\href {\doibase 10.1103/PhysRevLett.99.067003} {\bibfield
  {journal} {\bibinfo  {journal} {Phys. Rev. Lett.}\ }\textbf {\bibinfo
  {volume} {99}},\ \bibinfo {pages} {067003} (\bibinfo {year}
  {2007})}\BibitemShut {NoStop}%
\bibitem [{\citenamefont {Aji}\ and\ \citenamefont {Varma}(2009)}]{AjiVarma09}%
  \BibitemOpen
  \bibfield  {author} {\bibinfo {author} {\bibfnamefont {V.}~\bibnamefont
  {Aji}}\ and\ \bibinfo {author} {\bibfnamefont {C.~M.}\ \bibnamefont
  {Varma}},\ }\href {\doibase 10.1103/PhysRevB.79.184501} {\bibfield  {journal}
  {\bibinfo  {journal} {Phys. Rev. B}\ }\textbf {\bibinfo {volume} {79}},\
  \bibinfo {pages} {184501} (\bibinfo {year} {2009})}\BibitemShut {NoStop}%
\bibitem [{\citenamefont {Chang}\ \emph {et~al.}(2008)\citenamefont {Chang},
  \citenamefont {Wang},\ and\ \citenamefont {Zheng}}]{ChangWangZheng08}%
  \BibitemOpen
  \bibfield  {author} {\bibinfo {author} {\bibfnamefont {Z.}~\bibnamefont
  {Chang}}, \bibinfo {author} {\bibfnamefont {P.}~\bibnamefont {Wang}}, \ and\
  \bibinfo {author} {\bibfnamefont {Y.-H.}\ \bibnamefont {Zheng}},\ }\href
  {\doibase 10.1088/0253-6102/49/2/57} {\bibfield  {journal} {\bibinfo
  {journal} {Commun. Theor. Phys.}\ }\textbf {\bibinfo {volume} {49}},\
  \bibinfo {pages} {525} (\bibinfo {year} {2008})}\BibitemShut {NoStop}%
\bibitem [{\citenamefont {Ceccatto}(1991)}]{Ceccatto91}%
  \BibitemOpen
  \bibfield  {author} {\bibinfo {author} {\bibfnamefont {H.}~\bibnamefont
  {Ceccatto}},\ }\href@noop {} {\bibfield  {journal} {\bibinfo  {journal} {J.
  Phys. A}\ }\textbf {\bibinfo {volume} {24}},\ \bibinfo {pages} {2829}
  (\bibinfo {year} {1991})}\BibitemShut {NoStop}%
\bibitem [{\citenamefont {Baxter}(1982)}]{Baxter_book82}%
  \BibitemOpen
  \bibfield  {author} {\bibinfo {author} {\bibfnamefont {R.}~\bibnamefont
  {Baxter}},\ }\href@noop {} {\emph {\bibinfo {title} {Exactly Solved Models in
  Statistical Mechanics}}}\ (\bibinfo  {publisher} {Academic Press},\ \bibinfo
  {address} {New York},\ \bibinfo {year} {1982})\BibitemShut {NoStop}%
\bibitem [{\citenamefont {Hrahsheh}\ \emph {et~al.}(2014)\citenamefont
  {Hrahsheh}, \citenamefont {Hoyos}, \citenamefont {Narayanan},\ and\
  \citenamefont {Vojta}}]{HHNV14}%
  \BibitemOpen
  \bibfield  {author} {\bibinfo {author} {\bibfnamefont {F.}~\bibnamefont
  {Hrahsheh}}, \bibinfo {author} {\bibfnamefont {J.~A.}\ \bibnamefont {Hoyos}},
  \bibinfo {author} {\bibfnamefont {R.}~\bibnamefont {Narayanan}}, \ and\
  \bibinfo {author} {\bibfnamefont {T.}~\bibnamefont {Vojta}},\ }\href
  {\doibase 10.1103/PhysRevB.89.014401} {\bibfield  {journal} {\bibinfo
  {journal} {Phys. Rev. B}\ }\textbf {\bibinfo {volume} {89}},\ \bibinfo
  {pages} {014401} (\bibinfo {year} {2014})}\BibitemShut {NoStop}%
\bibitem [{\citenamefont {Sachdev}(1999)}]{Sachdev_book99}%
  \BibitemOpen
  \bibfield  {author} {\bibinfo {author} {\bibfnamefont {S.}~\bibnamefont
  {Sachdev}},\ }\href@noop {} {\emph {\bibinfo {title} {Quantum phase
  transitions}}}\ (\bibinfo  {publisher} {Cambridge University Press},\
  \bibinfo {address} {Cambridge},\ \bibinfo {year} {1999})\BibitemShut
  {NoStop}%
\bibitem [{\citenamefont {Wiseman}\ and\ \citenamefont
  {Domany}(1995)}]{WisemanDomany95}%
  \BibitemOpen
  \bibfield  {author} {\bibinfo {author} {\bibfnamefont {S.}~\bibnamefont
  {Wiseman}}\ and\ \bibinfo {author} {\bibfnamefont {E.}~\bibnamefont
  {Domany}},\ }\href {\doibase 10.1103/PhysRevE.51.3074} {\bibfield  {journal}
  {\bibinfo  {journal} {Phys. Rev. E}\ }\textbf {\bibinfo {volume} {51}},\
  \bibinfo {pages} {3074} (\bibinfo {year} {1995})}\BibitemShut {NoStop}%
\bibitem [{\citenamefont {Swendsen}\ and\ \citenamefont
  {Wang}(1987)}]{SwendsenWang87}%
  \BibitemOpen
  \bibfield  {author} {\bibinfo {author} {\bibfnamefont {R.~H.}\ \bibnamefont
  {Swendsen}}\ and\ \bibinfo {author} {\bibfnamefont {J.-S.}\ \bibnamefont
  {Wang}},\ }\href {\doibase 10.1103/PhysRevLett.58.86} {\bibfield  {journal}
  {\bibinfo  {journal} {Phys. Rev. Lett.}\ }\textbf {\bibinfo {volume} {58}},\
  \bibinfo {pages} {86} (\bibinfo {year} {1987})}\BibitemShut {NoStop}%
\bibitem [{\citenamefont {Wolff}(1989)}]{Wolff89}%
  \BibitemOpen
  \bibfield  {author} {\bibinfo {author} {\bibfnamefont {U.}~\bibnamefont
  {Wolff}},\ }\href@noop {} {\bibfield  {journal} {\bibinfo  {journal} {Phys.
  Rev. Lett.}\ }\textbf {\bibinfo {volume} {62}},\ \bibinfo {pages} {361}
  (\bibinfo {year} {1989})}\BibitemShut {NoStop}%
\bibitem [{Note1()}]{Note1}%
  \BibitemOpen
  \bibinfo {note} {Generalizations of the Swendsen-Wang and Wolff algorithms
  exist for systems with competing interaction, but they turn out be much less
  efficient \cite {KesslerBretz90,Liang92}}\BibitemShut {NoStop}%
\bibitem [{\citenamefont {Metropolis}\ \emph {et~al.}(1953)\citenamefont
  {Metropolis}, \citenamefont {Rosenbluth}, \citenamefont {Rosenbluth},\ and\
  \citenamefont {Teller}}]{MRRT53}%
  \BibitemOpen
  \bibfield  {author} {\bibinfo {author} {\bibfnamefont {N.}~\bibnamefont
  {Metropolis}}, \bibinfo {author} {\bibfnamefont {A.}~\bibnamefont
  {Rosenbluth}}, \bibinfo {author} {\bibfnamefont {M.}~\bibnamefont
  {Rosenbluth}}, \ and\ \bibinfo {author} {\bibfnamefont {A.}~\bibnamefont
  {Teller}},\ }\href@noop {} {\bibfield  {journal} {\bibinfo  {journal} {J.
  Chem. Phys.}\ }\textbf {\bibinfo {volume} {21}},\ \bibinfo {pages} {1087}
  (\bibinfo {year} {1953})}\BibitemShut {NoStop}%
\bibitem [{\citenamefont {Wang}\ and\ \citenamefont
  {Landau}(2001)}]{WangLandau01}%
  \BibitemOpen
  \bibfield  {author} {\bibinfo {author} {\bibfnamefont {F.}~\bibnamefont
  {Wang}}\ and\ \bibinfo {author} {\bibfnamefont {D.~P.}\ \bibnamefont
  {Landau}},\ }\href {\doibase 10.1103/PhysRevLett.86.2050} {\bibfield
  {journal} {\bibinfo  {journal} {Phys. Rev. Lett.}\ }\textbf {\bibinfo
  {volume} {86}},\ \bibinfo {pages} {2050} (\bibinfo {year}
  {2001})}\BibitemShut {NoStop}%
\bibitem [{\citenamefont {Zhu}\ \emph {et~al.}(2015)\citenamefont {Zhu},
  \citenamefont {Wan}, \citenamefont {Narayanan}, \citenamefont {Hoyos},\ and\
  \citenamefont {Vojta}}]{ZWNHV15}%
  \BibitemOpen
  \bibfield  {author} {\bibinfo {author} {\bibfnamefont {Q.}~\bibnamefont
  {Zhu}}, \bibinfo {author} {\bibfnamefont {X.}~\bibnamefont {Wan}}, \bibinfo
  {author} {\bibfnamefont {R.}~\bibnamefont {Narayanan}}, \bibinfo {author}
  {\bibfnamefont {J.~A.}\ \bibnamefont {Hoyos}}, \ and\ \bibinfo {author}
  {\bibfnamefont {T.}~\bibnamefont {Vojta}},\ }\href {\doibase
  10.1103/PhysRevB.91.224201} {\bibfield  {journal} {\bibinfo  {journal} {Phys.
  Rev. B}\ }\textbf {\bibinfo {volume} {91}},\ \bibinfo {pages} {224201}
  (\bibinfo {year} {2015})}\BibitemShut {NoStop}%
\bibitem [{\citenamefont {Guo}\ \emph {et~al.}(1994)\citenamefont {Guo},
  \citenamefont {Bhatt},\ and\ \citenamefont {Huse}}]{GuoBhattHuse94}%
  \BibitemOpen
  \bibfield  {author} {\bibinfo {author} {\bibfnamefont {M.}~\bibnamefont
  {Guo}}, \bibinfo {author} {\bibfnamefont {R.~N.}\ \bibnamefont {Bhatt}}, \
  and\ \bibinfo {author} {\bibfnamefont {D.~A.}\ \bibnamefont {Huse}},\
  }\href@noop {} {\bibfield  {journal} {\bibinfo  {journal} {Phys. Rev. Lett.}\
  }\textbf {\bibinfo {volume} {72}},\ \bibinfo {pages} {4137} (\bibinfo {year}
  {1994})}\BibitemShut {NoStop}%
\bibitem [{\citenamefont {Rieger}\ and\ \citenamefont
  {Young}(1994)}]{RiegerYoung94}%
  \BibitemOpen
  \bibfield  {author} {\bibinfo {author} {\bibfnamefont {H.}~\bibnamefont
  {Rieger}}\ and\ \bibinfo {author} {\bibfnamefont {A.~P.}\ \bibnamefont
  {Young}},\ }\href@noop {} {\bibfield  {journal} {\bibinfo  {journal} {Phys.
  Rev. Lett.}\ }\textbf {\bibinfo {volume} {72}},\ \bibinfo {pages} {4141}
  (\bibinfo {year} {1994})}\BibitemShut {NoStop}%
\bibitem [{\citenamefont {Sknepnek}\ \emph {et~al.}(2004)\citenamefont
  {Sknepnek}, \citenamefont {Vojta},\ and\ \citenamefont
  {Vojta}}]{SknepnekVojtaVojta04}%
  \BibitemOpen
  \bibfield  {author} {\bibinfo {author} {\bibfnamefont {R.}~\bibnamefont
  {Sknepnek}}, \bibinfo {author} {\bibfnamefont {T.}~\bibnamefont {Vojta}}, \
  and\ \bibinfo {author} {\bibfnamefont {M.}~\bibnamefont {Vojta}},\
  }\href@noop {} {\bibfield  {journal} {\bibinfo  {journal} {Phys. Rev. Lett.}\
  }\textbf {\bibinfo {volume} {93}},\ \bibinfo {pages} {097201} (\bibinfo
  {year} {2004})}\BibitemShut {NoStop}%
\bibitem [{\citenamefont {Vojta}\ and\ \citenamefont
  {Sknepnek}(2006)}]{VojtaSknepnek06}%
  \BibitemOpen
  \bibfield  {author} {\bibinfo {author} {\bibfnamefont {T.}~\bibnamefont
  {Vojta}}\ and\ \bibinfo {author} {\bibfnamefont {R.}~\bibnamefont
  {Sknepnek}},\ }\href@noop {} {\bibfield  {journal} {\bibinfo  {journal}
  {Phys. Rev. B.}\ }\textbf {\bibinfo {volume} {74}},\ \bibinfo {pages}
  {094415} (\bibinfo {year} {2006})}\BibitemShut {NoStop}%
\bibitem [{\citenamefont {Vojta}\ \emph {et~al.}(2016)\citenamefont {Vojta},
  \citenamefont {Crewse}, \citenamefont {Puschmann}, \citenamefont {Arovas},\
  and\ \citenamefont {Kiselev}}]{Vojtaetal16}%
  \BibitemOpen
  \bibfield  {author} {\bibinfo {author} {\bibfnamefont {T.}~\bibnamefont
  {Vojta}}, \bibinfo {author} {\bibfnamefont {J.}~\bibnamefont {Crewse}},
  \bibinfo {author} {\bibfnamefont {M.}~\bibnamefont {Puschmann}}, \bibinfo
  {author} {\bibfnamefont {D.}~\bibnamefont {Arovas}}, \ and\ \bibinfo {author}
  {\bibfnamefont {Y.}~\bibnamefont {Kiselev}},\ }\href {\doibase
  10.1103/PhysRevB.94.134501} {\bibfield  {journal} {\bibinfo  {journal} {Phys.
  Rev. B}\ }\textbf {\bibinfo {volume} {94}},\ \bibinfo {pages} {134501}
  (\bibinfo {year} {2016})}\BibitemShut {NoStop}%
\bibitem [{\citenamefont {Barber}(1983)}]{Barber_review83}%
  \BibitemOpen
  \bibfield  {author} {\bibinfo {author} {\bibfnamefont {M.~N.}\ \bibnamefont
  {Barber}},\ }in\ \href@noop {} {\emph {\bibinfo {booktitle} {Phase
  Transitions and Critical Phenomena}}},\ Vol.~\bibinfo {volume} {8},\ \bibinfo
  {editor} {edited by\ \bibinfo {editor} {\bibfnamefont {C.}~\bibnamefont
  {Domb}}\ and\ \bibinfo {editor} {\bibfnamefont {J.~L.}\ \bibnamefont
  {Lebowitz}}}\ (\bibinfo  {publisher} {Academic},\ \bibinfo {address} {New
  York},\ \bibinfo {year} {1983})\ pp.\ \bibinfo {pages} {145--266}\BibitemShut
  {NoStop}%
\bibitem [{\citenamefont {Cardy}(1988)}]{Cardy_book88}%
  \BibitemOpen
  \bibinfo {editor} {\bibfnamefont {J.}~\bibnamefont {Cardy}},\ ed.,\
  \href@noop {} {\emph {\bibinfo {title} {Finite-size scaling}}}\ (\bibinfo
  {publisher} {North Holland},\ \bibinfo {address} {Amsterdam},\ \bibinfo
  {year} {1988})\BibitemShut {NoStop}%
\bibitem [{\citenamefont {Binder}(1983)}]{Binder83}%
  \BibitemOpen
  \bibfield  {author} {\bibinfo {author} {\bibfnamefont {K.}~\bibnamefont
  {Binder}},\ }\href@noop {} {\bibfield  {journal} {\bibinfo  {journal} {Z.
  Phys. B}\ }\textbf {\bibinfo {volume} {50}},\ \bibinfo {pages} {343}
  (\bibinfo {year} {1983})}\BibitemShut {NoStop}%
\bibitem [{\citenamefont {Grinstein}\ and\ \citenamefont
  {Ma}(1983)}]{GrinsteinMa83}%
  \BibitemOpen
  \bibfield  {author} {\bibinfo {author} {\bibfnamefont {G.}~\bibnamefont
  {Grinstein}}\ and\ \bibinfo {author} {\bibfnamefont {S.-k.}\ \bibnamefont
  {Ma}},\ }\href {\doibase 10.1103/PhysRevB.28.2588} {\bibfield  {journal}
  {\bibinfo  {journal} {Phys. Rev. B}\ }\textbf {\bibinfo {volume} {28}},\
  \bibinfo {pages} {2588} (\bibinfo {year} {1983})}\BibitemShut {NoStop}%
\bibitem [{\citenamefont {Kessler}\ and\ \citenamefont
  {Bretz}(1990)}]{KesslerBretz90}%
  \BibitemOpen
  \bibfield  {author} {\bibinfo {author} {\bibfnamefont {D.~A.}\ \bibnamefont
  {Kessler}}\ and\ \bibinfo {author} {\bibfnamefont {M.}~\bibnamefont
  {Bretz}},\ }\href {\doibase 10.1103/PhysRevB.41.4778} {\bibfield  {journal}
  {\bibinfo  {journal} {Phys. Rev. B}\ }\textbf {\bibinfo {volume} {41}},\
  \bibinfo {pages} {4778} (\bibinfo {year} {1990})}\BibitemShut {NoStop}%
\bibitem [{\citenamefont {Liang}(1992)}]{Liang92}%
  \BibitemOpen
  \bibfield  {author} {\bibinfo {author} {\bibfnamefont {S.}~\bibnamefont
  {Liang}},\ }\href {\doibase 10.1103/PhysRevLett.69.2145} {\bibfield
  {journal} {\bibinfo  {journal} {Phys. Rev. Lett.}\ }\textbf {\bibinfo
  {volume} {69}},\ \bibinfo {pages} {2145} (\bibinfo {year}
  {1992})}\BibitemShut {NoStop}%
\end{thebibliography}%

\end{document}